\documentclass[aps,pra,superscriptaddress,twocolumn,twocolumn,10pt]{revtex4-1}
\usepackage{amsmath}
\usepackage{latexsym}
\usepackage{graphicx}
\usepackage{amsfonts}
\usepackage{braket}
\usepackage{hyperref}
\usepackage{natbib}
\usepackage{amssymb}
\usepackage{nicefrac}
\usepackage{fancyhdr}
\usepackage[utf8]{inputenc}
\usepackage{dsfont}
\usepackage{colortbl}
\usepackage{xcolor}
\usepackage{subfigure}
\usepackage{psfrag}
\usepackage{tikz}
\usepackage{paralist}
\usepackage{nicefrac}
\usepackage{ulem}
\usepackage{placeins}
\usepackage{mathtools}
\usepackage{bbm}

\usetikzlibrary{arrows,shapes,decorations.pathmorphing}

\makeindex

\begin{document}

\title{Flux Qubit Readout in the Persistent Current Basis at arbitrary Bias Points}

\author{M. Sch\"ondorf}
\affiliation{Theoretical Physics, Saarland University, 66123
Saarbr{\"u}cken, Germany}
\author{A. Lupa\c{s}cu}
\affiliation{Institute for Quantum Computing, Department of Physics and Astronomy, and Waterloo Institute for Nanotechnology, University of Waterloo, Waterloo, Ontario, Canada N2L 3G1}
\author{F. K. Wilhelm}
\affiliation{Theoretical Physics, Saarland University, 66123
Saarbr{\"u}cken, Germany}

\begin{abstract}

Common flux qubit readout schemes are qubit dominated, meaning they measure in the energy eigenbasis of the qubit. For various applications, measurements in a basis different from the energy eigenbasis are required. Here we present an indirect measurement protocol, which is detector dominated instead of qubit dominated, yielding a projective measurements in the persistent current basis for arbitrary bias points. We show, that with our setup it is possible to perform a quantum nondemolition measurement in the persistent current basis at all flux bias points with fidelities reaching almost $100\%$. 
\end{abstract}

\maketitle

\section{Introduction}

The measurement postulate is fundamental in the formulation of quantum mechanics \cite{NielsenChuan}. To obtain information about the quantum state of a closed system one needs to employ an interaction with an additional readout system (meter). It is possible to design this interaction such that the measured observable is an integral of motion during the readout process. This is called a quantum non-demolition (QND) measurement. QND measurements enable repeated measurements to have the same outcome and were originally proposed to exceed the standard quantum limit in connection with the detection of gravitational waves \cite{braginsky1980quantum,braginsky1996quantum,danilishin2012quantum}. The interest in QND measurement methods has increased with the development of quantum information, where they play an important role in various aspects, e.g. error correction \cite{fowler2012surface} or initialization by measurement \cite{ruskov2003entanglement}.

Superconducting flux qubits \cite{mooij1999josephson} are especially interesting for the field of quantum annealing \cite{farhi2001quantum,jansen2007bounds,aharonov2008adiabatic,bravyi2008quantum,amin2009consistency,RevModPhys.90.015002,PhysRevB.81.134510,PhysRevApplied.8.014004}, where the intrinsic possibility for inductive coupling and the rather large anharmonicity deliver a big advantage. However, for flux qubits QND measurements in the persistent current basis have only been performed far away from the flux degeneracy point \cite{lupacscu2005quantum,lupacscu2006high,boulant2007quantum,mallet2009single,PhysRevLett.93.177006}. At the degeneracy point the expectation value of the persistent current, which is the measurement variable, is zero for the qubit energy eigenstates. Measurement in the energy eigenbasis at the degeneracy point is possible by coupling the qubit transversely to a resonator, leading to a measurement of the quantum inductance \cite{orgiazzi2016flux,inomata2012large,rifkin1976current,greenberg2002low}, or by using a more complicated scheme based on modulated coupling \cite{wang2011ideal}. The ability to perform measurements in the flux basis at an arbitrary operation point is especially interesting in quantum annealing. To be able to measure during the anneal process without first driving the qubit far away from the degeneracy point would yield huge advantages, e.g. avoid quenches in anealing schedules, which limit success probability \cite{RevModPhys.90.015002,harris2018phase,king2018observation} or realize quantum speedup with only stoquastic interactions \cite{fujii2018quantum}. In addition, state tomography would benefit from such a readout scheme, since measurements in canonical conjugated basis are necessary \cite{lvovsky2009continuous,filipp2009two}.
\\
\begin{figure}
\includegraphics[width=.3\textwidth]{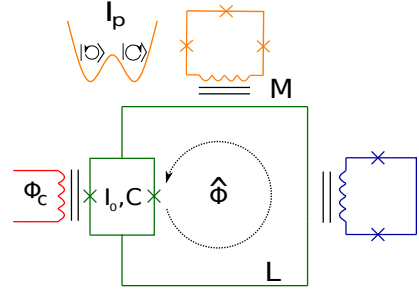}
\caption{Circuit for the measurement protocol. The qubit (yellow) is coupled to the large SQUID loop of the quantum probe, here the cjj-SQUID (green). The large loop is coupled to an additional flux readout loop (blue) and an external control flux $\Phi_c$ (red) is applied to the small SQUID loop.}
\label{system_setup}
\end{figure}
Here we present a method to measure the state of a flux qubit for arbitrary biases, ranging from the symmetry point to points far from the symmetry point, which is both projective and high fidelity. In contrast to usual flux qubit measurements \cite{lupacscu2005quantum,lupacscu2006high,boulant2007quantum,mallet2009single,PhysRevLett.93.177006}, we measure in the persistent current basis at all bias points and not in the energy eigenbasis of the qubit. 

The paper is organized as follows. In Sec. \ref{sec:2} we present our setup the corresponding Hamiltonian. The four different steps of the measurement protocol are discussed in detail in Sec \ref{sec:3}. In Sec. \ref{sec:4} we present our results and in Sec. \ref{sec:5} we give the conclusion.

\section{Setup and Hamiltonian}

\label{sec:2}

The proposed indirect measurement protocol includes a quantum probe in between the flux qubit we want to read out and the actual readout resonator (e.g. SQUID). This probe is a compound Josephson junction SQUID (cjj-SQUID) \cite{harris2009compound,poletto2009tunable,castellano2010deep}. The cjj-SQUID is coupled inductively to the superconducting flux qubit we want to measure, leading to the Hamiltonian for the coupled qubit-probe system (for setup see Fig. \ref{system_setup})
\begin{align}
\hat H = \frac{\phi_0^2}{L}\left(4\xi^2 \frac{\hat q^2}{2} + \frac{\hat \varphi^2}{2}+\beta_{\rm cjj}(\Phi_c)\cos\hat\varphi- \frac{g}{\sqrt{\xi}}\hat \varphi \hat \sigma_z\right) + \hat H_{\rm qb},
\label{Hamiltonian}
\end{align}
where $\xi = e/\phi_0\sqrt{L/C_{\Sigma}}$, $\phi_0 = \Phi_0/2\pi$ and $g= \sqrt{\xi} M I_p/\phi_0$, with mutual inductance $M$, persistent current $I_p$, inductance of the large cjj-SQUID loop $L$, sum of the two junction capacitances $C_{\Sigma} $ and $\Phi_0$ the flux quantum. The quantum variable of the probe is the average phase of the junctions $\hat \varphi = 2\pi\hat\Phi/\Phi_0$ and $\hat q$ is the conjugated variable.  $\hat H_{qb}$ denotes the flux qubit Hamiltonian represented in the persistent current basis $ \{\ket{\circlearrowleft},\ket{\circlearrowright}\}$
\begin{align}
\hat H_{\rm qb} = \frac{\epsilon}{2}\hat\sigma_z + \frac{\Delta}{2}\hat \sigma_x,
\end{align} 
with energy spacing $\epsilon$ and tunneling energy $\Delta$. Note that we do not include the Hamiltonian of the readout loop in Eq. \eqref{Hamiltonian}, since it is decoupled during the whole dynamics of interest and only used after the protocol is performed to readout the persistent current state of the probe. 

A special property of the cjj-SQUID is that the screening parameter depends on the additional control flux $\Phi_c$ applied to the small loop, i.e.  $\beta_{cjj}(\Phi_c) = (2I_0 L/\phi_0) \cos(\Phi_c/2\phi_0)$ \cite{castellano2010deep}, with critical current of the SQUID junctions $I_0$. The measurement starts with the cjj-SQUID operated in a regime where the potential is parabolic, and centered at a value that depends on the state of the qubit. Next, the control flux $\Phi_c$ is used to transform the potential into a double well barrier potential, leading to states localized in one of the wells, in correspondence with the two qubit states. In contrast to usual measurement schemes, here we present a detector dominated measurement by choosing strong or even ultrastrong coupling \cite{kockum2019ultrastrong} between the qubit and the quantum probe, such that the measured observable is determined by the eigenbasis of the operator coupled to the probe. Here this is the persistent current basis, as opposed to the qubit energy eigenbasis. We show that our measurement protocol enables an almost perfect QND measurement at the degeneracy point and can achieve measurement fidelities close to $100\%$. Note that the protocol also works in a similar way for  $\epsilon \neq 0$.

\section{Measurement protocol}
\label{sec:3}

\begin{figure*}[t!]
\includegraphics[width=.8\textwidth]{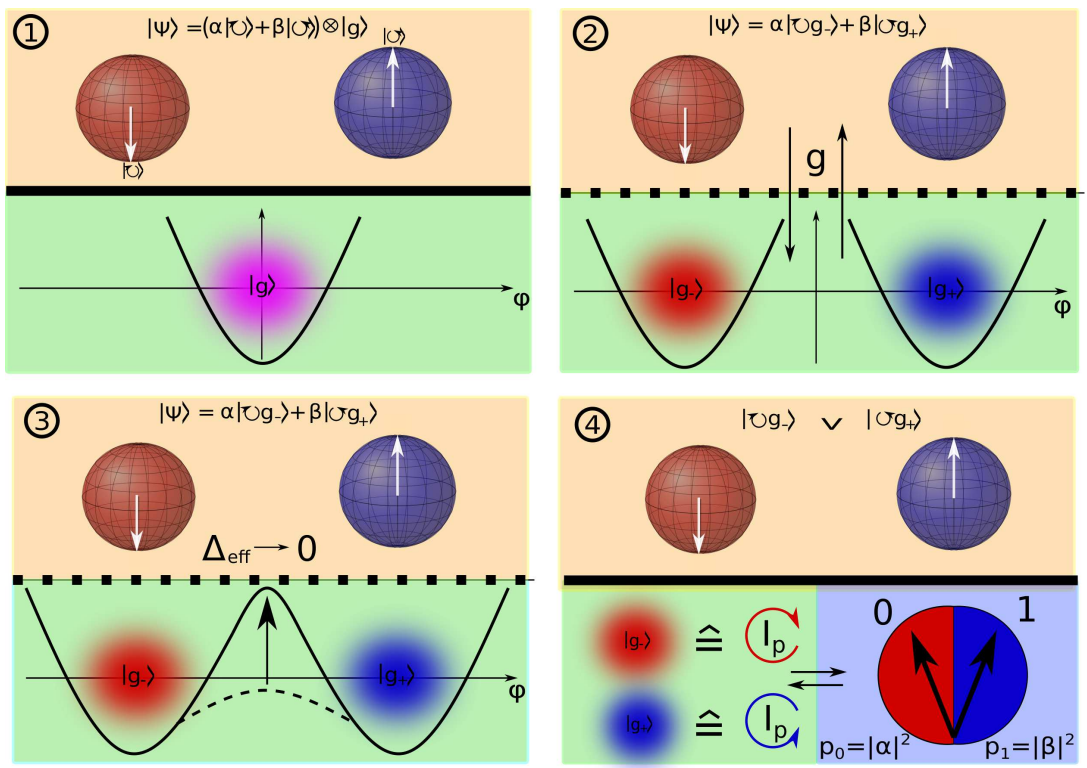}
\caption{Principle of the measurement scheme. Color code analog to Fig \ref{system_setup}. 1.) Initialization: The qubit (yellow) and the cjj-SQUID (green) initial state are prepared. 2.) Premeasurement: The coupling between the qubit and the cjj-SQUID is ramped up, such that the qubit states get entangled with corresponding pointer states. 3.) Effective decoupling: The c-jj SQUID potential is turned from a single well to a double well potential, resulting in an exponential decrease of the effective coupling $\Delta_{\rm eff}$. 4.) Readout: The cjj-SQUID persistent current state is read out with an additional flux readout device (e.g rf-SQUID).}
\label{measurement_protocol}
\end{figure*}

The measurement protocol is schematically shown in Fig. \ref{measurement_protocol}. It has four main steps: the initialization, the premeasurement, the effective decoupling and the readout of the probe.

In the initialization step we prepare the qubit in an arbitrary initial state $\alpha \ket{\circlearrowleft}+\beta\ket{\circlearrowright}$ and the cjj-SQUID in the ground state $\ket{g}$. Here the qubit and the probe are decoupled and the screening parameter $\beta_{cjj}$ of the cjj-SQUID is zero, meaning it is described by a harmonic oscillator potential. 

After intialization we start the premeasurement. For this, we turn on the coupling between the qubit and the cjj-SQUID. During this step, the external bias on the small coupler loop is still choosen as $\Phi_c = \Phi_0/2$, such that the barrier is zero and the cjj-SQUID potential is purely quadratic. By turning on the coupling between the cjj-SQUID and the qubit, an entangled state between the qubit and the pointer states of the probe is created, performing the premeasurement. Since the cjj-SQUID starts in the ground state $\ket{g}$, the coupling term shifts the center of the Gaussian distribution of the phase. For zero coupling, the cjj-SQUID state is centered around $\left<\hat\varphi\right> = 0$, until the coupling shifts the mean value to $\left<\hat \varphi\right> =  \varphi_p \left<\hat \sigma_z\right>$, where $\varphi_p$ denotes the absolute value of the cjj-SQUID potential minimum. Note that the shift depends on the qubit state as follows from \eqref{Hamiltonian}.

Here we want to choose parameters such that the interaction does not induce any excitation of the cjj-SQUIDs initial ground state, meaning we require a perfect adiabatic time evolution of the system \cite{wilhelm2003asymptotic}
\begin{align}
\left(\alpha \ket{\circlearrowleft} + \beta \ket{\circlearrowright} \right)\ket{g} \longrightarrow \alpha_{\rm eff} \ket{\circlearrowleft,g_-} + \beta_{\rm eff} \ket{\circlearrowright,g_+},
\end{align}
where $\ket{g}$ is the coupler ground state centered around zero and $\ket{g_{\pm}}$ are the corresponding displaced ground states.  $\alpha_{\rm eff}$ and $\beta_{\rm eff}$ include the time evolution under the bare qubit Hamiltonian \cite{wilhelm2003asymptotic}.There are two factors that change the probability amplitudes $\alpha$ and $\beta$. On the one hand if the state is not an eigenstate of $\hat H_{\rm qb}$, it evolves under the bare qubit Hamiltonian. However as we will see later, this is strongly depressed by the ramping of the barrier. On the other hand the coupling to the quantum probe leads to a measurement induced dephasing, meaning the phase information encoded in $\alpha$ and $\beta$ gets lost during the measurement, as we will also see in more detail in the next section (for more details see App. \ref{app:2}). The effective coupling energy $\Delta_{\rm eff}$ gets rescaled due to the interaction with the cjj-SQUID \cite{wilhelm2003asymptotic}. To make the adiabatic approximation applicable, the timescale of the interaction must satisfy the adiabatic theorem \cite{Kato1950}. This yields the condition
\begin{align}
{\rm max}_{t} \frac{\dot g(t)}{\sqrt{\xi}} \ll \Omega,
\label{adiabaticity}
\end{align}
with characteristic frequency of the quadratic part of the cjj-SQUID, $\Omega =  1/\sqrt{LC}$. Violating this condition leads to transitions between cjj-SQUID pointer states which destroys the distinguishability, since there is no longer a clear map between direction of persistent current and qubit state.

Besides the fact that we want the measurement to discriminate between the qubit states, we additionally want the measurement to be QND. A QND measurement is achieved, when the measured observable is an integral of motion during the measurement, meaning successive measurements of the qubit yield the same result \cite{braginsky1992measurement}. This is achieved by the third step of our protocol, the effective decoupling. Especially in the case $\epsilon \ll \Delta$, the non-commuting part of the system and the interaction Hamiltonian is crucial, hence severe backaction would appear during the macroscopic readout of the probe. Therefore in the effective decoupling step, we use the external bias $\Phi_c$ to ramp the barrier of the cjj-SQUID potential from a single well harmonic potential to a double well potential with a high barrier. This exponentially decreases the effective coupling energy $\Delta_{\rm eff}$, resulting in a reduction of the non-commuting part. Here $\Delta_{\rm eff}$ means the tunneling matrix element between the two compound states $\ket{\circlearrowright,g_-}$ and $\ket{\circlearrowleft,g_+}$. With this we freeze the dynamics of the qubit, yielding an effective decoupling of the qubit and the probe, necessary for a QND measurement \cite{braginsky1992measurement}. Note that the tuning of the barrier also has to be adiabatically on the cjj-SQUID timescale to again avoid excitations to higher modes, such that we have to modify condition \eqref{adiabaticity} and include the time derivative of the screening parameter $\beta_{cjj}(t)$
\begin{align}
{\rm max}_{t}\left[ \frac{\dot g(t)}{\sqrt{\xi}} , \dot \beta_{cjj}(t)\right] \ll \Omega.
\end{align}
 
In a last step we can measure the probe state using the additional persistent current readout with indicating almost no backaction, since $\Delta_{\rm eff}(T) \approx 0$, where $T$ denotes the time for the overall protocol.

Because of the non-commuting nature of the interaction and the system Hamiltonian, there is also a back action induced during the premeasurement and the effective decoupling. Therefore one needs to perform these two steps fast with respect to the characteristic qubit timescale
\begin{align}
 T\sqrt{\Delta^2+\epsilon^2} \ll h,
\end{align}
where $h$ is the Planck constant. However, this general condition is too strict in our case. On the one hand the whole point of the third step is to decrease the effective decoupling rate to almost zero and on the other hand in the case $\epsilon \gg \Delta$ the backaction is negligible, since system and interaction Hamiltonian almost commute. Including these facts, the QND condition for our system is given by 
\begin{align}
\int_0^T \Delta_{\rm eff}(t) {\rm d}t \ll h.
\label{condition}
\end{align}
Note that the effective tunneling rate is time dependent, since it is influenced by the interaction with the probe. Because of the entanglement of the pointer states and the qubit states after the premeasurement, a high barrier of the cjj-SQUID potential also frustrates a tunneling between the qubit states. This leads to the fact that \eqref{condition} is even satisfied for measurement times larger than the qubits characteristic time, as we will see in the next section.

At the end of the decoupling step, it is important that the two pointer states are statistically distinguishable, meaning that the maximal coupling strength  $g_{\rm max} = g(T)$, needs to be chosen such that the condition \cite{clerk2003quantum}
\begin{align}
\left<\varphi(T)\right>_{1} - \left<\varphi(T)\right>_{0} \geq 2 \left[\sigma_1(T) + \sigma_0(T)\right],
\label{distinguishability}
\end{align}
is satisfied at the end. Condition $\eqref{distinguishability}$ is a qualitative measure for statistical distinguishability, but does not quantify measurement fidelity. Here $\left<\varphi(T)\right>_{i}$ is the expectation value of the pointer state if the qubit is in state $i$ and $\sigma(T)_i$ is the respective standard deviation. Both are taken at the end of the measurement protocol. 
\begin{figure}
\includegraphics[width=.4\textwidth]{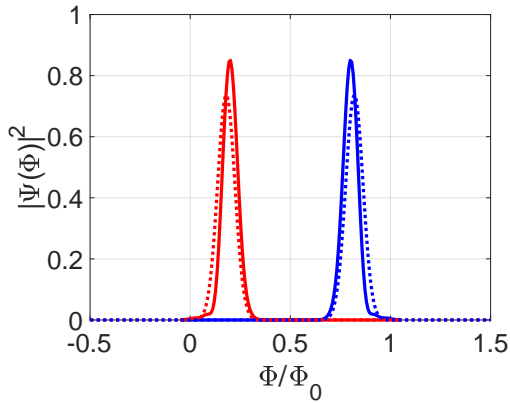}
\caption{Results of the numerical simulations of the measurement protocol for $\Delta/h = 0.1\Omega$ and linear time schedule with $g_{\rm max}/\Omega = 1$ and $\beta_{cjj}^{\rm max} = 2$ and $\xi = 0.1$ (corresponding e.g. to $C_{\Sigma} = 15$ fF and $L=600$ pH). The initial qubit state is chosen to be $\ket{0}=1/\sqrt{2}(\ket{\circlearrowleft}+\ket{\circlearrowright})$. Shown is the probability distribution of projection to qubit left (red) and right (blue) persistent current state. Numerical (solid) and analytical results (dotted).}
\label{results}
\end{figure}

The distinguishability criterion gives a lower bound for the necessary maximal coupling strength $g_{\rm max}$. The measurement fidelity is limited by the overlap of the pointer states (see Fig . \ref{results}) and transitions between different cjj-SQUID states during the interaction process. Therefore the most general expression for the measurement fidelity is given by 
\begin{align}
\mathcal{F}_{\rm meas} &= \frac{\mathcal{F}_{\circlearrowright}+\mathcal{F}_{\circlearrowleft}}{2},
\label{fidelity}
\end{align}
where $\mathcal{F}_i$ denotes the probability to get the right measurement result if the qubit is prepared in the energy eigenstate $\ket{i}$. The state fidelities read
\begin{align}
\mathcal{F}_{i} &= \int_{a_i}^{b_i}\left(\left|\braket{\varphi,\circlearrowleft|\hat U(t)|g,i}\right|^2 + \left|\braket{\varphi,\circlearrowright|\hat U(t)|g,i}\right|^2\right) {\rm d}\varphi  \label{state_fidelity1} 
\end{align}
where $i \in \{\circlearrowleft,\circlearrowright\}$, $\hat U(t)$ is the time evolution operator which describes the time dynamics of the measurement process and $\{a_{\circlearrowleft},b_{\circlearrowleft}\} = \{-\infty,0\}$, $\{a_{\circlearrowright},b_{\circlearrowright}\} = \{0,\infty\}$. 

\begin{figure}
\includegraphics[width=.4\textwidth]{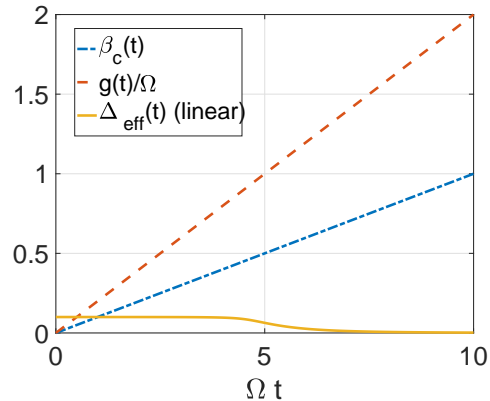}
\caption{Time evolution of the screening parameter $\beta_{\rm cjj}$, the coupling $g$ and the coupling energy $\Delta_{\rm eff}$.}
\label{parameters}
\end{figure}

\section{Results}

\label{sec:4}

We want to quantitatively study the measurement protocol. The most important point here is to quantify the right time scales and system parameters to obtain high measurement fidelities and prove the QNDness of the protocol. To solve the time dependent Hamiltonian \eqref{Hamiltonian} numerically, we evolve the cjj-SQUID part in harmonic oscillator modes. Here we truncate the Hamiltonian after $100$ excitations. Since $g(t)$ and $\beta_{\rm cjj}(t)$ are time dependent we have to solve a time dependent Schrödinger equation. For this we use a standard Runge-Kutta method. In the simulations we assume that the coupling and the nonlinearity are turned on simultaneously instead of successively. We simulate the full Hamiltonian \eqref{Hamiltonian} including the qubit dynamics. 

We choose the simplest possible time dependence here, where we tune up the coupling and the barrier linearly. Here the maximal value of the coupling is $g_{\rm max}/\Omega=1$ and the maximal screening parameter $\beta_{cjj}^{\rm max}$ is $2$. The overall time interval in which we ramp up both parameters is chosen to be $10/\Omega$ and $\Omega$ is ten times the qubit frequency. The time evolution of the coupling, the screening parameter and the linear approximation of the effective coupling strength (see App. \ref{app:2} for more details on how to calculate this) are shown in Fig. \ref{parameters}. We study the measurement protocol at the flux degeneracy point $\epsilon = 0$ and choose as initial state the qubit eigenstate $\ket{0} = (\ket{\circlearrowleft}+\ket{\circlearrowright})/\sqrt{2}$.
\begin{figure}
\includegraphics[width=.4\textwidth]{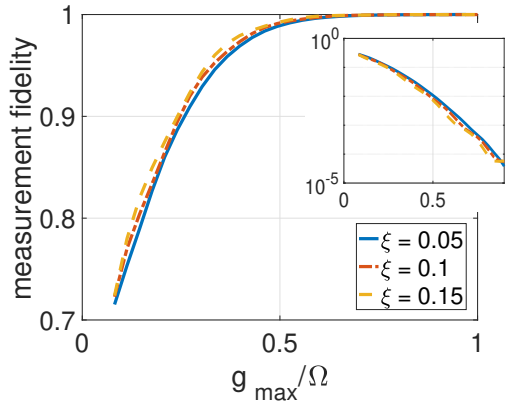}
\caption{Measurement fidelity depending on $g_{\rm max}$ for different values of $m$ for the same parameters as in Fig. \ref{results} and for different values of $\xi$. The inset shows the infidelity in log scale to illustrate how the fidelity tends to one for large couplings.}
\label{fidelity_plot}
\end{figure}
After modeling the time evolution of the pointer states using \eqref{Hamiltonian}, we calculate the measurement fidelity given by Eq. \eqref{fidelity}. Fig. \ref{results} proves that the pointer states of the cjj-SQUID nicely resolve the qubit states. The measurement error is given by the overlap of the two probability distributions. For the chosen parameters this results in a measurement fidelity of $1$, since the overlap of the states is zero.

To quantify this, we also show the measurement fidelity dependence on the maximal coupling strength $g_{\rm max}$ for different values of $\xi$ in Fig \ref{fidelity_plot}. We see that the measurement fidelity strongly increases for larger values of $g_{\rm max}$ until it reaches a plateau at fidelity 1. For smaller $\xi$, the fidelities are slightly lower, but do not vary significantly in the range of realistic system parameters \cite{PhysRevLett.104.177004,PhysRevLett.118.057702,neill2016ergodic,castellano2010deep}. Even though the ultrastrong coupling regime is accessible in flux qubit architectures (\cite{niemczyk2010circuit,baust2016ultrastrong,forn2017ultrastrong}), it is more feasible to work in the strong coupling regime. However, even in this regime which corresponds to $g_{\rm max}/\Omega \approx 0.1$, the measurement fidelities are quite high. E.g. for $\xi=0.1$ we reach a fidelity of $80.8\%$ for the same measurement time as in Fig. \ref{results}. Since the coupling is weaker the interaction time needed for a resolving premeasurement is also longer, hence it is supporting to choose longer measurement times. For an increased measurement time of $T =40/\Omega$ the fidelity for $\xi =0.1$ and $g=0.1$ already reaches $95\%$ and can be further improved by decreasing $T$. This means that there is a tradeoff between coupling and measurement time. One can choose a smaller coupling when at the same hand the measurement time is increased, but the coupling is roughly lower bounded by condition \eqref{distinguishability}. Note that a measurement time of $10/\Omega$ in our choice of parameters corresponds to the characteristic time scale of the qubit. For flux qubits this would be in the ns regime. 

Here we model measurement at the flux degeneracy point, but the protocol can lead to perfect fidelities  for $\epsilon\neq 0$, e.g. for the same parameters as in Fig. \ref{results} but for the case $\epsilon/h = \Delta/h = \Omega/10$, our simulations show an almost perfect measurement fidelity of $0.999$. The simulations are performed in exactly the same way as for the case $\epsilon = 0$.

In Fig. \ref{results}, also the time evolution of the density matrix elements for the initial qubit state $\ket{0}=1/\sqrt{2}(\ket{\circlearrowleft}+\ket{\circlearrowright})$ at the degeneracy point is studied. The parameters are the same as before. We see that the measurement induces a strong dephasing in the measurement basis (persistent current basis). This is what one expects since entangling the qubit with the respective pointer states means transferring qubit information to the probe system (\cite{breuer2002theory} or \cite{braginsky1992measurement}). The fact that the measurement induces a dephasing in the persistent current basis proves that the meausrement protocol does not measure in the energy eigenbasis of the qubit, but in the eigenbasis of the probe. The diagonal elements on the other hand stay constant, meaning the population in the persistent current basis is conserved.
\begin{figure}
\includegraphics[width=.4\textwidth]{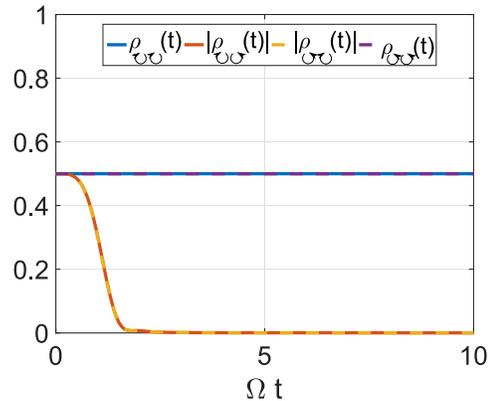}
\caption{Decay of excited state population during measurement for initial state $\ket{\circlearrowleft}$ and the same parameters as in Fig \ref{results}.}
\label{QND1}
\end{figure}

As mentioned before, a way to determine the QNDness of a measurement is the comparison of repeated successive readouts. Since here the measurement observable is the persistent current, we have to study the decay of the corresponding states $\ket{\circlearrowright}, \ket{\circlearrowleft}$ of the qubit to check for QNDness. The QNDness in our system can be quantified as the probability that the qubits initial persistent current state is preserved after premeasurement, irrespective of the measurement outcome \cite{braginsky1992measurement,lupacscu2007quantum}, yielding the expression
\begin{align}
\mathcal{F}_{\rm QND} = \frac{\braket{\circlearrowright|\hat U_{\rm QB}(t)|\circlearrowright}+\braket{\circlearrowleft|\hat U_{\rm QB}(t)|\circlearrowleft}}{2},
\label{QND}
\end{align}
where $\hat U_{\rm QB}(t) = {\rm Tr}_{\rm probe}\{\hat U(t)\}$ denotes the effective time evolution of the qubit during the premeasurement. This means nothing else than successive measurements giving the same results which is the textbook definition of QNDness. As mentioned, to ensure QNDness of the protocol we ramp up the barrier and effectively discriminate the time evolution of the system, which leads to the fact that only in the beginning of the protocol the qubit suffers a small rotation. Fig. \ref{results} shows the decay of the diagonal density matrix element during the measurement protocol, which is in the order of $10^{-3}$ for the chosen parameters. Expression \eqref{QND} can be determined numerically and yields a QND fidelity of $\mathcal{F}_{\rm QND} = 99.6\%$.  We are optimistic that further optimization strategies (e.g. find optimized schedules) could lead to even better results, yielding a perfect QND measurement with a measurement fidelity of $100\%$. 
\begin{figure}
\includegraphics[width=.4\textwidth]{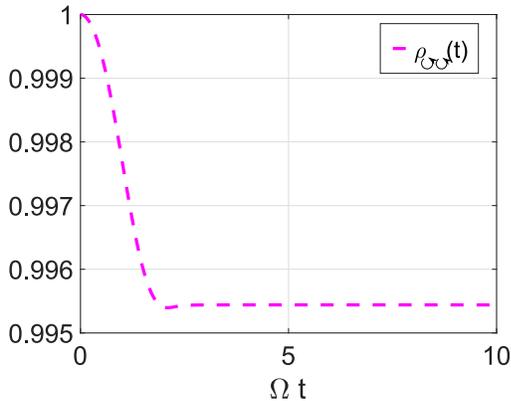}
\caption{Decay of excited state population during measurement for initial state $\ket{\circlearrowleft}$ for the same parameters as in Fig. \ref{results}.}
\label{QND2}
\end{figure}

The quantum probe, here realized by the cjj-SQUID includes an additional device compared to conventional flux qubit schemes. Such an additional device could make the system more susceptible to noise. In flux qubit architectures, the most present noise is $1/f$-flux noise \cite{PhysRevB.76.064531,PhysRevLett.98.267003,PhysRevLett.97.167001}. In App. \ref{app_noise} we model the influence of $1/f$-noise acting on the cjj-SQUID of our measurement circuit. It can be seen that neither the measurement fidelity is changed significantly (error in the order of $10^{-3}$) due to the noise, nor additional back action is induced on the qubit. This shows that the presented protocol is not more susceptible to noise than conventional flux qubit readout architectures.  

Using a Gaussian approximation for the pointer states and assuming a completely adiabatic time evolution, it is also possible to derive an analyitcal expression for the probabiltiy amplitude of the pointer states. A comparison between analytical and numerical results can be found in Fig. \ref{results}. We see that both results match very well, allthough there is a small deviation arising from higher order potential terms, i.e. the numerical distributions are slightly asymmetric and shifted towards $\Phi/\Phi_0 = 0.5$. The measurement fidelity within the Gaussian approximation is given by 
\begin{align}
\mathcal{F}_{\rm meas} = \Phi\left(\frac{\varphi_p(T)}{\sigma(T)}\right),
\end{align} 
where $\varphi_p$ is the position of the minima of the cjj-SQUID double well potential at the end of the protocol, $\sigma(T) = (2m\Omega\sqrt{1-\beta_{cjj}(T)\cos\varphi_p(T)})^{-1/2}$ is the standard deviation and $\Phi(x)$ denotes the normal cumulative distribution function. The detailed calculations can be found in the Supplement, where we additionally study the backaction analytically and show a qualitative agreement with the numerically found back action results. 

For the sense of completeness, we want to point out that D-wave also uses the cjj-SQUID as a qubit but that the measurement method differs from the one presented here. They use a quench to first tune the qubits into the regime $\epsilon \gg \Delta$ and then perform a persistent current readout (see e.g. \cite{harris2018phase}). 

\section{Conclusion}

\label{sec:5}

In conclusion we have presented an indirect measurement protocol to perform fast read out a flux qubit at every bias point in the persistent current basis, with possible measurement fidelities close to $100\%$. Further the measurement is also shown to be QND, which increases the possibility for applications in fundamental flux qubit experiments as well as in the perspective of quantum annealing even more. A special feature is that the readout at the flux degeneracy point is performed in the persistent current basis, being potentially useful in terms of quantum annealing but also for other applications such as quantum state tomography.

\section*{Acknowledgments}

This material is based upon work supported by the Intelligence Advanced Research Projects Activity (IARPA) and the Army Research Office (ARO) under Contract No. W911NF-17-C-0050. Any opinions, findings and conclusions or recommendations expressed in this material are those of the authors and do not necessarily reflect the views of the Intelligence Advanced Research Projects Activity (IARPA) and the Army Research Office (ASO). Further we thank Simon Jäger for fruitful discussions.

\begin{appendix}
\section{Analytical Results}
\label{app1}

\subsection{Measurement fidelity}
\label{app:1}

In this section we will analytically describe the setup presented in Sec I, especially giving approximate expression for the success probability.

Since the regime of interest is $\Omega \gg \epsilon,\Delta$, we consider the qubit Hamiltonian as the perturbation of the system
\begin{align}
V &= \frac{\epsilon}{2}\sigma_z + \frac{\Delta}{2}\sigma_x.
\end{align}
As shown in the main part, the phase-charge space representation of the unperturbed Hamiltonian reads
\begin{align}
H_0(t) = \frac{q^2}{2m} + m\Omega^2 \frac{\varphi^2}{2} + m\Omega^2\beta_c(t) \cos(\varphi) - m\Omega^2\lambda(t) \sigma_z \varphi.
\end{align}
with effective mass $L/(2\xi\phi_0)^2$. Without the cosine term, this yields a shifted harmonic oscillator where the shift depends on the qubit state. To include the contribution of the non-harmonic cosine part, we will approximate the potential around its minimum. It is
\begin{align}
U'(\varphi) = m\Omega^2\varphi - m\Omega^2\beta_c(t) \sin(\varphi) - m\Omega^2\lambda(t)\sigma_z.
\end{align}
The condition $U'(\varphi) = 0$ leads to an equation for the potential minimum, depending on $\sigma_z$. Since $\sigma_z = \pm 1$ the position is symmetric for the two qubit states
\begin{align}
\varphi_{\pm}(t) = \pm \varphi_p(t),
\label{phi_p}
\end{align}
where $\varphi_p(t)$ denotes the positive valued minimum.
The effective potential up to second order then reads
\begin{align}
U(\varphi) &\approx \varphi_p(t)\sigma_z + \frac{m\Omega^2}{2} \left[1-\beta_c(t)\cos(\varphi_p(t))\right] \left(\varphi - \varphi_p(t)\hat \sigma_z \right)^2 \\
&= \varphi_p(t) \sigma_z + \frac{m\tilde\Omega(t)^2}{2} \varphi^2 - m\tilde\Omega(t)^2 \varphi_p(t) \varphi \hat \sigma_z.
\end{align}
with time dependent frequency $\tilde\Omega(t) = \Omega\sqrt{1-\beta_c(t)\cos(\varphi_{p})}$. Note that the frequency does not depend on the qubit state, because of the symmetry of the cosine. This leads to the effective Hamiltonian
\begin{align}
\hat H_0(t)&\approx  \frac{q^2}{2m} + \frac{m \tilde \Omega(t)^2 \varphi^2}{2} - m\tilde\Omega(t)^2\varphi_{p}(t) \varphi\hat\sigma_z \\
&= \tilde\Omega(t) a^{\dag}a - \tilde\Omega(t)\sqrt{\frac{m\tilde\Omega(t)}{2}} \varphi_p(t) (a^{\dag}+a)\hat\sigma_z.
\end{align}
The last part implies a qubit dependent shift of the harmonic oscillator, such that we can diagonalize this Hamiltonian with the displacement operator
\begin{align}
\hat{\tilde H}(t) &= D^{\dag}(\tilde \varphi_p(t)\hat\sigma_z) H D(\tilde \varphi_p(t)\hat\sigma_z) \\
&= \tilde \Omega(t) a^{\dag}a\label{Hamilton_trafo}
\end{align}
where $\tilde \varphi_p(t) = \varphi_p(t) \sqrt{m\tilde\Omega(t)/2}$. The time dependence of the transformation induces an additional inertia term. As mentioned before, we choose time scales to be diabatic on the qubit and adiabatic on the coupler time scale. Hence in zeroth order we assume the SQUID state to follow the minimum adiabatically, so we ignore the term proportional to $\dot{\tilde \varphi}_p$ (inertia part) for now. Additionally we ignore the contribution arising from the zeroth order of the Taylor expansion, since it only acts as a correction of the bare qubit Hamiltonian (for more details see App. \ref{app:2}). 

We can directly write down the solution to \eqref{Hamilton_trafo} in the position space which is a Gaussian distribution around the minimum of the potential
\begin{align}
\varphi(t) = \left(2\pi \sigma(t)^2\right)^{1/4} {\rm e}^{\left(\frac{\varphi-\varphi_p\left<\hat\sigma_z\right>(t)}{2\sigma(t)}\right)^2 + ip_0 \varphi} \ket{\varphi},
\end{align}
with standard deviation $\sigma(t) = 1/\sqrt{2 m \tilde\Omega(t)}$ and $p_0$ being the average momentum. Let us now assume the qubit starts in a superposition state and the cjj-SQUID in its ground state (centered around $\varphi = 0$). The time evolution reads
\begin{align}
\left(\alpha \ket{\circlearrowleft} + \beta \ket{\circlearrowright}\right)\ket{g} \overset{\hat U}{\rightarrow} \alpha_{\rm eff} \ket{\circlearrowleft,\varphi_-(t)} + \beta_{\rm eff} \ket{\circlearrowright,\varphi_+(t)}, 
\end{align}
with
\begin{align}
\ket{\varphi_\pm(t)} = \left(2\pi\sigma(t)^2\right)^{-1/4} {\rm e}^{\left(\frac{\varphi\mp\varphi_p}{2\sigma(t)}\right)^2+ip_0 \varphi} \ket{\varphi} 
\end{align}
and where $\alpha_{\rm eff}$ and $\beta_{\rm eff}$ include the time evolution induced by the bare qubit Hamiltonian, i.e when the state is not an eigenstate (see \cite{wilhelm2003asymptotic} for more details).
We are especially interested in the probabilities for the SQUID to be in the left or right persistent current state, depending on the qubit state. E.g. the probability to get the right measurement result if the qubit starts in the $\ket{\circlearrowleft}$ state (equivalent to $\mathcal{F}_{\circlearrowleft}$ of the main text) is given by 
\begin{align}
\mathcal{F}_{\circlearrowleft}(t) &= \frac{1}{\sqrt{2\pi \sigma(t)^2}}  \int_{-\infty}^0 {\rm e}^{\frac{\left(\varphi+\varphi_p\right)^2}{2\sigma(t)^2}} {\rm d}\varphi\\
&= \Phi\left(\frac{\varphi_p(T)}{\sigma(t)}\right),  
\end{align}
with $\Phi(x) = \frac{1}{\sqrt{2\pi}} \int_{-\infty}^x {\rm e}^{- \frac{1}{2}t^2} {\rm d}t$ denoting the normal cumulative distribution function. In the same manner we can write down the probability to get the right measurement result when the qubit starts in state $\ket{1}$
\begin{align}
\mathcal{F}_{\circlearrowright}(T) = -\Phi\left(-\frac{\varphi_p(T)}{\sigma(T)}\right).
\end{align}
This expressions correspond to the two contributions that appear in the expression for the fidelity, hence in the Gaussian approximation $\mathcal{F}$ can be written as
\begin{align}
\mathcal{F}(T) = \Phi\left(\frac{\varphi_p(T)}{\sigma(T)}\right),
\end{align}
where we used the fact that $\Phi(t)$ is an odd function. Fortunately, Gaussians are among the simplest special functions and the expectation value is completely determined by the standard deviation $\sigma(T)$, hence the fidelity is fully determined by $\sigma(T)$ and $\varphi_p$. This fact can be used to e.g. put a lower bound on the measurement fidelity and determine the corresponding system parameter intervals to reach this fidelity. Here the main parameters that can be varied are $\lambda_{\rm max}$ and $\beta_{\rm cjj}^{\rm max}$. One could also optimize the schedule, i.e. find an optimal pulse for the time dynamics of the coupling and the barrier to optimize both, measurement fidelity and back action. However, this would yield an optimal control problem and can be tracked by future work. A lower bound for the respective system parameters is given by the distinguishability condition (Eq. (3)). Since the distributions are symmetric, the condition has the simplified form
\begin{align}
\varphi_p(T) \geq 2 \sigma(T) 
\end{align}  

\begin{figure}
\includegraphics[width=.4\textwidth]{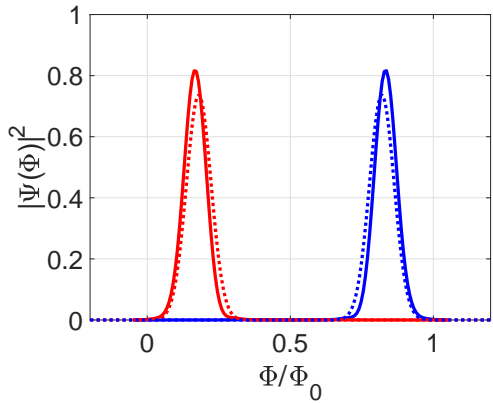}
\caption{Comparison of the numerical (solid) and analytical (dotted) results for the same parameters as in the main text.}
\label{prob_plus_an}
\end{figure}

In Fig. \ref{prob_plus_an} the distribution of the cjj-SQUID state (depending on the qubit state) is compared to the numerical results. We basically see what we expect; the two results qualitatively coincide but there are corrections coming from the higher order potential terms. Since we model the double well potential of the cjj-SQUID with two harmonic potentials, the two actual expectation are slightly shifted compared to the Gaussian ones. Additionally the width of the actual distribution is also slightly smaller.

In App. \ref{app:2} we use the same strategy to calculate an expression for the time evolution of the density matrix in the cjj-SQUID ground subspace. Since the calculations are rather involved we put them into the appendix. The analytics show the right qualitative and long time behavior but differ quantitatively from the numerical results caused by different approximations made during the calculation.

All in all this section shows, that the intuitive picture of the system dynamics, we gave when we described the measurement scheme in the main text can be quantified with the given analytical results assuming an adiabatic time evolution of the pointer states. Since the analytical results also give a good agreement with the numerics, the adiabatic approximation is satisfied for the chosen time scale, avoiding any induced transitions between different cjj-SQUID states. Further the given results could be used to optimize system parameters for real world applications.

\subsection{Backaction}

\label{app:2}
Here we will try to analytically approximate the back action of the measurement on the qubit. For this we first transform the Hamiltonian into an interaction frame (i.e the displaced oscillator frame) such that we can write down the time dependent Hamiltonian as a tensor sum of two dimensional matrices (within the adiabatic approximation). Then we can study the time evolution of the qubit subspace density matrix and with this make statements about the back action. 

As shown in \eqref{app:1} we can diagonalize $H_0$ approximately by applying the displacement operator
\begin{align}
\hat D(\tilde \varphi_p(t)\hat \sigma_z).
\end{align} 
This leads to a diagonal Hamiltonian plus an additional inertia term coming from the time dependence of the transformation and a correction of the bare qubit Hamiltonian arising from the fact that the two minima of the tilted double well potential are not at the same potential level 
\begin{align}
\tilde H_0 &= \tilde \Omega(t) \hat a^{\dag}a - i\dot{\tilde\varphi}_p(t)(a^{\dag}-a) -\lambda \varphi(t) \hat\sigma_z\\
&= \tilde \Omega(t) \hat a^{\dag}a - i\dot{\tilde\varphi}(t)p_0\left(\dot \varphi_p+\frac{1}{4}\frac{\dot{\tilde\Omega}(t)}{\tilde\Omega(t)}\varphi_p(t)\right) -\lambda \varphi(t) \hat\sigma_z
\end{align}
where $p_0(t)$ is the average momentum at time $t$, which can be rewritten using the correspondence principle $p_0(t) = m \dot \varphi_p(t)$. The last term arises from the zeroth order of the Taylor expansion. Hence we need to take into account two correction terms. We also have to check what is the effect of the transformation on the bare qubit Hamiltonian
\begin{align}
\tilde V &= \hat D^{\dag}(\tilde \varphi_p(t)\hat \sigma_z)\left[ \epsilon \hat \sigma_z + \Delta \hat \sigma_x\right]  \hat D(\tilde \varphi_p(t)\hat \sigma_z)\\
&\approx \epsilon \hat \sigma_z + \Delta \hat \sigma_x + 2 \tilde \varphi_p(t) p_0(t) \Delta \hat \sigma_y,
\end{align}
where we only kept the first order term of the Baker-Campbell-Hausdorff formula. Since we assume $\Delta \ll \Omega$, the $\sigma_y$ correction is assumed to be rather small compared to the $\sigma_z$ correction arising from $\hat{\tilde H}_0$, hence will be ignored in the following. With this we can write the Hamiltonian in the transformed basis as a tensor sum
\begin{align}
\tilde H(t) = \oplus_{N=0}^{\infty} \tilde H_N(t),
\end{align}
where $H_N(t)$ is the Hamiltonian in the $N$ excitation subspace $\{\ket{0,N_-},\ket{1,N_+}\}$ and has the form
\begin{align}
\tilde H_N(t) = \begin{pmatrix}
N\tilde\Omega(t) - \gamma(t) && \frac{\Delta}{2} \braket{N_+|N_-} \\ \frac{\Delta}{2} \braket{N_-|N_+} && N\tilde \Omega + \gamma(t)\end{pmatrix},
\end{align}
with
\begin{align}
\gamma(t) &= -m\left(\dot{\varphi}_p(t)^2+\frac{1}{4}\frac{\dot{\tilde\Omega}(t)}{\tilde\Omega(t)}\varphi_p(t)+\lambda\varphi_p(t)\right) \\
\tilde\Omega_N(t) &= N\tilde\Omega(t). 
\end{align}
Here $\ket{N_{\pm}}$ refer to the $N$ excitation states of a shifted harmonic oscillator, where the sign depends on the qubit state and the shift is given by \eqref{phi_p} (for more details on the shifted harmonic oscillator we refer to \cite{irish2005dynamics}). Here we assumed an adiabatic time evolution of the cjj-SQUID dynsmics by setting $\braket{N_{\pm}|M_{\pm}} = \braket{N_{\pm}|M_{\mp}}=0$ if $N\neq M$. Note that $\braket{N_+|N_-} = \braket{N_-|N_+}$, hence $\tilde H(t)$ is hermitian as demanded. The overlapp between the shifted oscillator vacuum states is given by \cite{irish2005dynamics}
\begin{align}
\braket{0_-|0_+} = {\rm e}^{-\varphi_p^2/2}
\end{align}
Because of the block diagonal structure of the Hamiltonian, we can also write down the time propagator $U(t)$ in a block diagonal structure. For this we need the following expressions
\begin{align}
U_N(t) = \exp\left(i \mathcal{T} \int_0^t {\rm d}t'\tilde H_N(t')\right),
\end{align}
with the time ordering operator $\mathcal{T}$. Since we assume the time evolution to be diabatic on the qubit subspace and we are interested in the dominating back action effects, we use first order Magnus expansion to calculate the time propagators $V_N(t)$
\begin{align}
V_N(t) \approx \exp\left(i\int_{0}^t {\rm d}t' H(t')\right). 
\end{align}
Defining the parameters $\Gamma(t) = \int_0^t {\rm d}t' \gamma(t')$ and $\tilde \Delta_N(t) = \Delta/2\int_0^t {\rm d}t' \braket{N_+|N_-}(t')$ the propagator of the $N$ excitation subspace can be written as
\begin{widetext}
\begin{tiny}
\begin{align}
V_N(t) = {\rm e}^{\int_0^t {\rm d}t'\tilde \Omega_N(t')}\begin{pmatrix} \cos\left(\sqrt{\Gamma(t)^2 + \tilde \Delta_N(t)^2}\right) - i \frac{\Gamma(t)}{\sqrt{\Gamma(t)^2+\tilde\Delta_N(t)^2}}\sin\left(\sqrt{\Gamma(t)^2+\tilde\Delta_N(t)^2}\right) && i\frac{\tilde \Delta_N}{\sqrt{\Gamma(t)^2+\tilde\Delta_N(t)^2}}\sin\left(\sqrt{\Gamma(t)^2+\tilde\Delta_N(t)^2}\right) \\ i\frac{\tilde \Delta_N}{\sqrt{\Gamma(t)^2+\tilde\Delta_N(t)^2}}\sin\left(\sqrt{\Gamma(t)^2+\tilde\Delta_N(t)^2}\right) && \cos\left(\sqrt{\Gamma(t)^2 + \tilde \Delta_N(t)^2}\right) + i \frac{\Gamma(t)}{\sqrt{\Gamma(t)^2+\tilde\Delta_N(t)^2}}\sin\left(\sqrt{\Gamma(t)^2+\tilde\Delta_N(t)^2}\right)\end{pmatrix}.
\end{align}
\end{tiny}
\end{widetext}
Since the back action tends to be strongest at the degeneracy point, we choose $\epsilon = 0$ in the following, such that $H_{\rm QB} = \frac{\Delta}{2}\sigma_x$. We want to study the time evolution of an arbitrary qubit state, when we prepare the SQUID in the ground state ($\left<N\right>=0$), leading to the following density matrix at $t=0$
\begin{align}
\hat \rho(0) = \begin{pmatrix}|\alpha|^2 && \alpha\beta^* \\ \alpha^*\beta && |\beta|^2\end{pmatrix} \otimes \ket{0}\bra{0}
\end{align}
The time evolution of this state can then be calculated using $V(t)$. We are especially interested in the density matrix of the qubit at time $t$, so we trace out the cjj-SQUID degrees of freedom
\begin{align}
\rho^{\rm QB}(t) &= {\rm Tr}_{\rm cjj}\left\{\rho(t)\right\} \\\begin{split}
&= |\alpha(t)|^2\ket{\circlearrowleft}\bra{\circlearrowleft} +\alpha(t)\beta^*(t)\ket{\circlearrowleft}\bra{
\circlearrowright}{\rm e}^{-\tilde \varphi_p(t)^2} \\&\hspace{0.4cm}  + \alpha^*(t)\beta(t)\ket{\circlearrowright}\bra{\circlearrowleft}{\rm e}^{-\tilde \varphi_p(t)^2} + |\beta(t)|^2 \ket{\circlearrowright}\bra{\circlearrowright}\end{split}.
\end{align}
Here we clearly see the measurement induced dephasing appearing as an exponential damping of the off diagonal elements, depending on the displacement between the two pointer states. The time evolution of the prefactors $\alpha$ and $\beta$ can be calculated using the time propagator. For the initial state $\ket{+}$ we have also chosen the main text, it is  $\alpha = 1/\sqrt{2}$ and $\beta = 1/\sqrt{2}$ leading to the density matrix entries
\begin{align}
\rho_{00}^{\rm QB}(t) &= \frac{1}{2}\left(1-2\frac{\tilde\Delta_0(t)\Gamma(t)}{\kappa(t)}\sin^2\kappa(t)\right)\\\begin{split}
\rho_{01}^{\rm QB}(t) &= \frac{1}{2}\left(1-2\frac{\Gamma(t)^2}{\kappa^2(t)}\sin^2\kappa(t)\right.\\ &\left.\hspace{0.6cm}-2i\frac{\Gamma(t)}{\kappa(t)}\sin\kappa(t)\cos\kappa(t)\right)\exp\left(-\tilde \varphi_p(t)^2\right)\end{split}\\
\rho_{10}^{\rm QB}(t) &= \left(\rho_{01}^{\rm QB}(t)\right)^* \\
\rho_{11}^{\rm QB}(t) &= 1-\rho_{00}^{\rm QB}(t),
\end{align} 
where we defined $\kappa(t) = \sqrt{\Delta_0(t)^2+\Gamma(t)^2}$. In Fig. \ref{parameters} we see the time evolution of the parameters $\Gamma(t)$ and $\tilde \Delta(t)$. We see that for $t\mapsto T$, $\Gamma$ gets much larger than $\Delta$ leading the oscillating term of the diagonal elements to go to zero, such that at the end of the measurement process the population is the same as in the beginning, proving the measurement to be QND. The long time behavior of the off diagonal elements are dominated by the measurement induced dephasing, i.e. the exponential part. Therefore the offdiagonal elements completely decay for $t\mapsto T$, what we also see in the numerical results. 
\begin{figure}
\includegraphics[width=.4\textwidth]{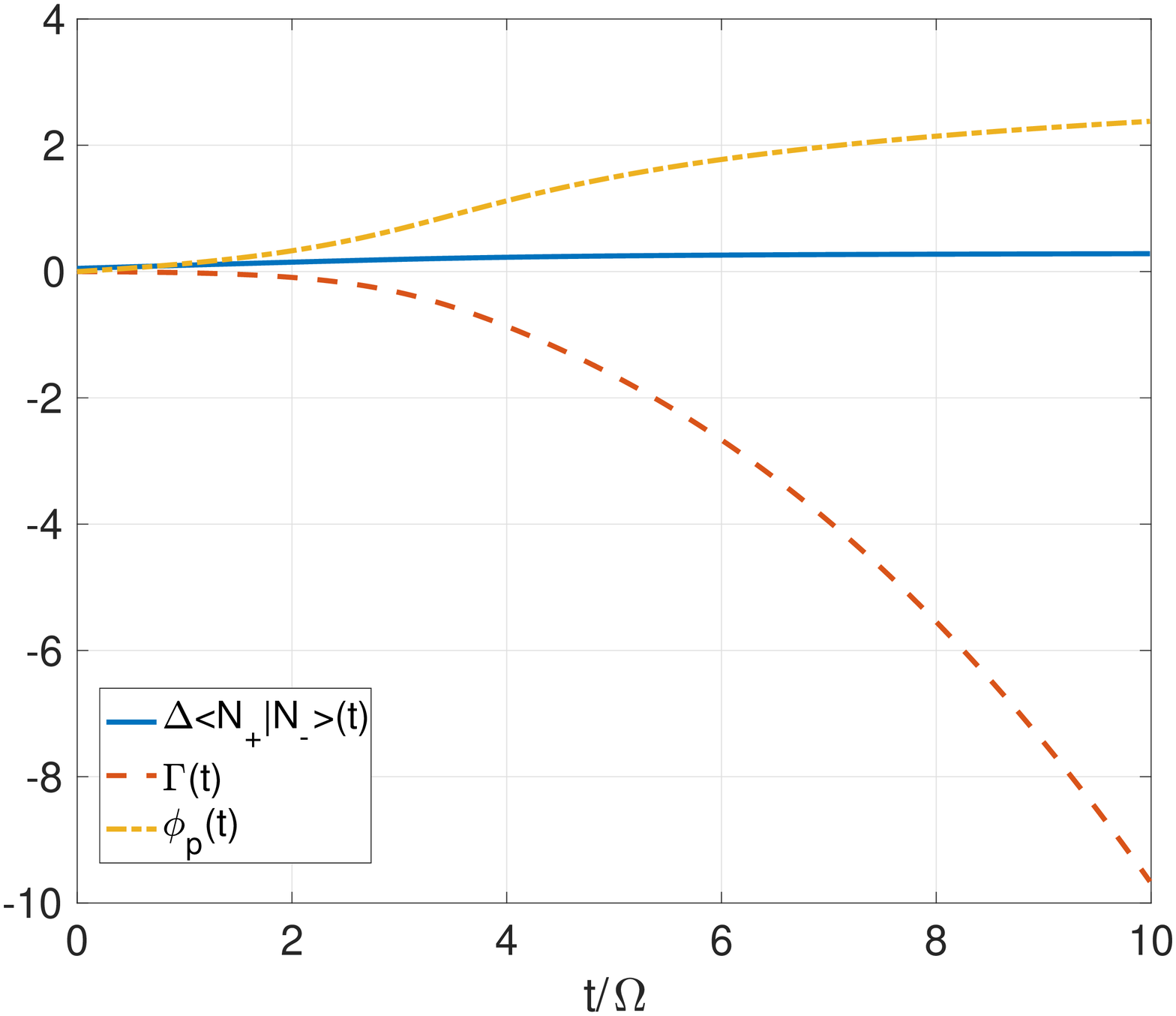}
\caption{Time evolution of the parameters $\tilde\Delta(t)$, $\Gamma(t)$ and $\varphi_p(t)$ for the same parameters as in the main text.}
\label{parameters2}
\end{figure}

However, even though the analytical results predict the right qualitative behavior and the right long time behavior, there are deviations between the analytical and numerical results. E.g. the predicted damped osicllations of the diagonal elements around $1/2$ are not observed in Fig. \ref{parameters2}. Two main factors limit the validity of the analytics. First we only included the first order of the Magnus expansion, but since $T$ is in the order of the qubit time evolution for the chosen parameters, it is not completely reasonable to assume a diabatic time evolution on the qubit time scale. Hence to get more rigorous results one has to include higher orders of the Magnus expansion. Second, we ignored the contributions coming from non commutating character of the interaction and the qubit Hamiltonian. Even though the studied contributions are the leading back action terms, for $T$ comparable to the qubit time scale, the other contributions also start to matter.
\begin{figure}
\includegraphics[width=.23\textwidth]{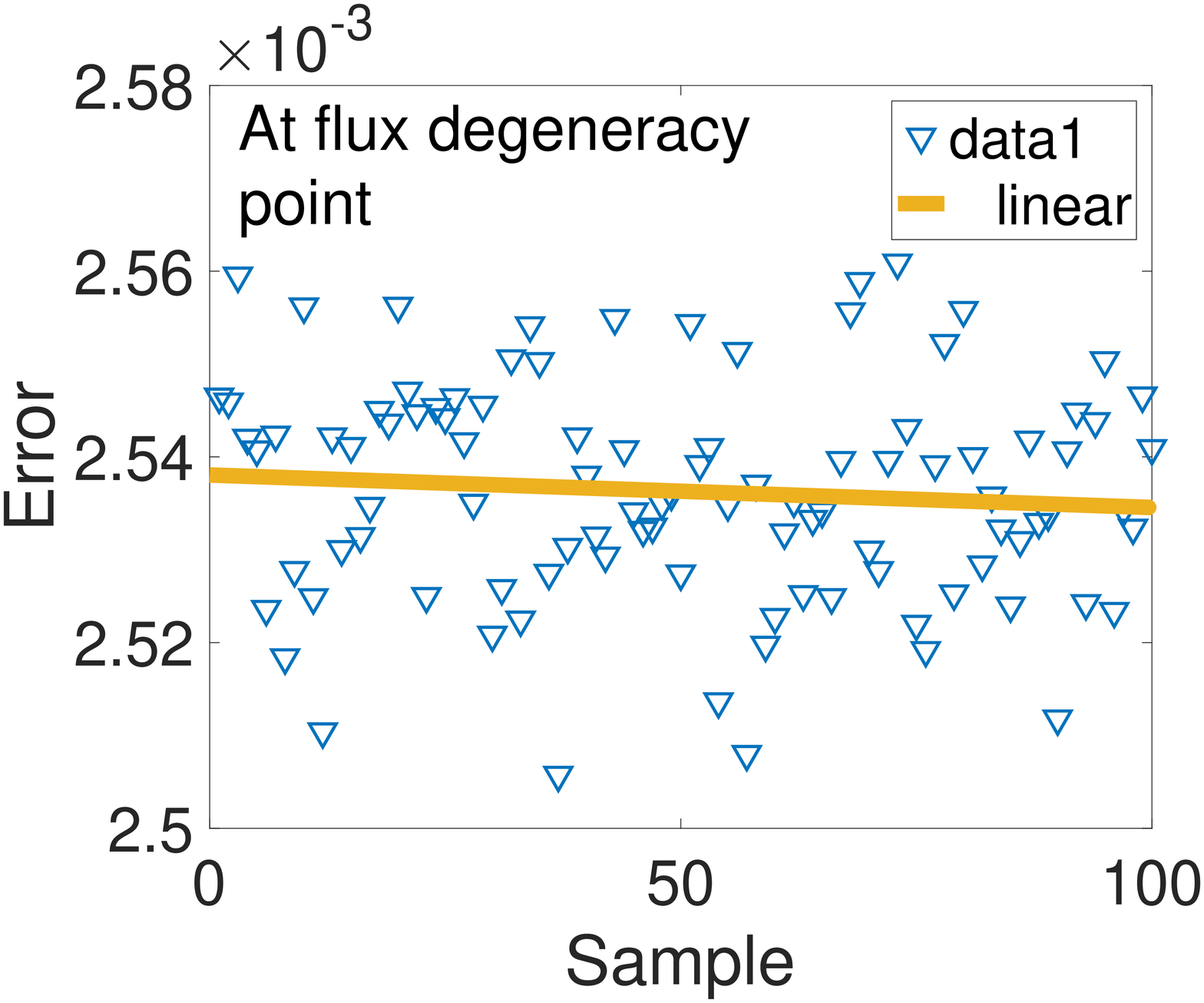}
\includegraphics[width=.23\textwidth]{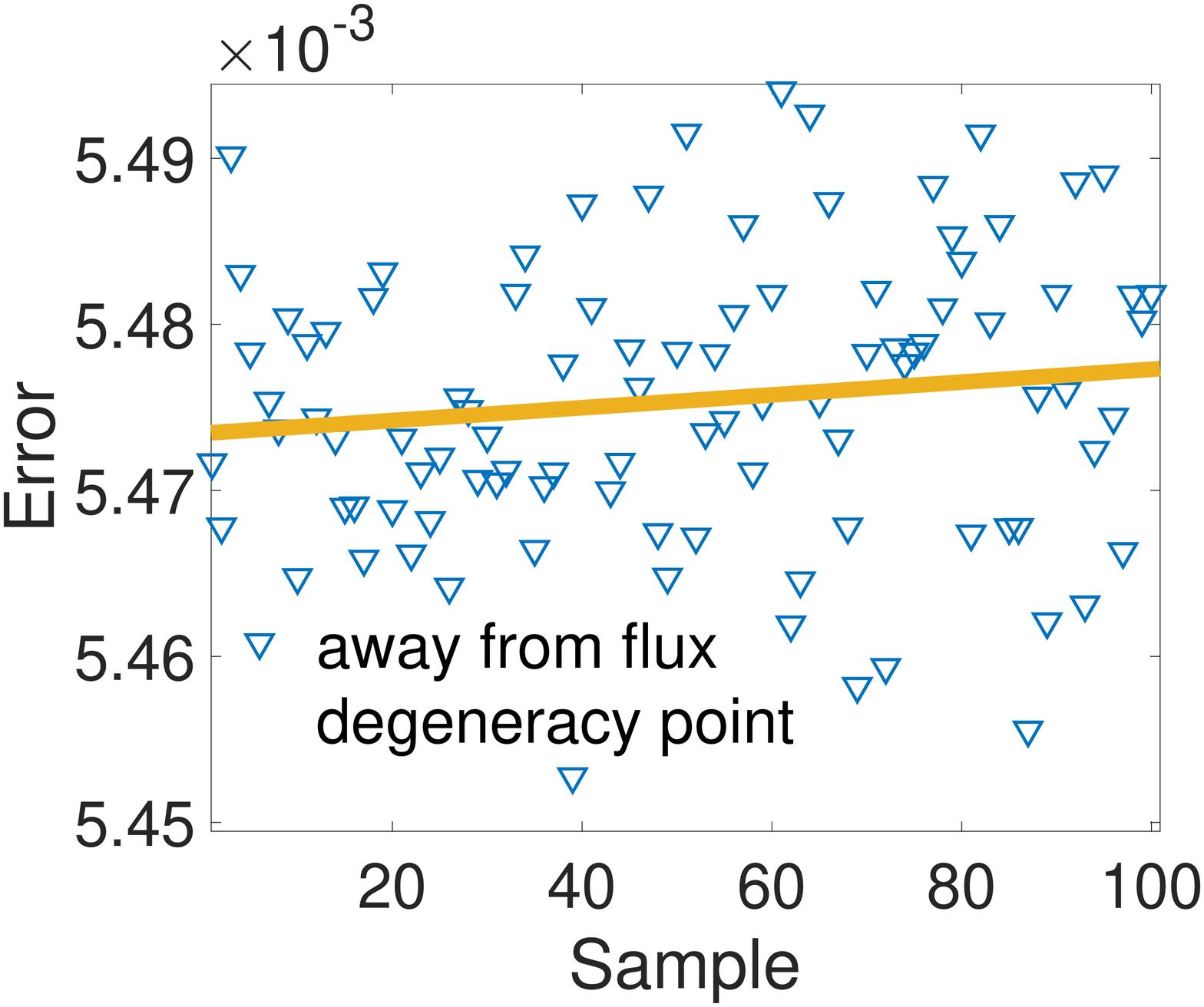}
\caption{Error of the measurement result at the flux degeneracy point(left) and way from it for $\epsilon=\Delta$ (right) and $100$ different generations of $1/f$-noise. The yellow line shows a linear fit to visualize the average error.}
\label{noise_cjj}
\end{figure}

\subsection{Flux noise acting on the cjj-SQUID}

\label{app_noise}

\begin{figure}
\includegraphics[width=.45\textwidth]{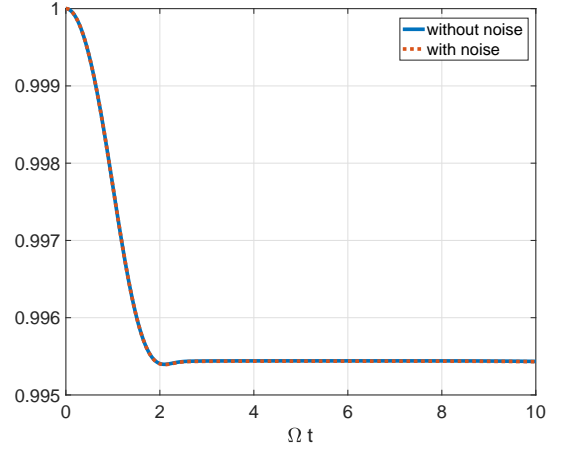}
\caption{Left:Evolution of the density matrix element $\rho_{\circlearrowleft \circlearrowleft}(t)$ for a vacuum environment (solid blue) and a $1/f$-noise environment (dotted yellow).}
\label{backaction_1f}
\end{figure}
Here we will study an effect which is more specific for our setup, i.e. flux noise acting on the quantum probe during the measurement. We assume $1/f$ flux noise, which typically appears in superconducting flux qubit architectures. To generate the noise trajectories, we use the matlab inbuilt object \textit{dsp.ColoredNoise}, which creates $1/|f|^{\alpha}$ noise, with $\alpha$ to choose from $[-2,2]$, using Gaussian sampling. Here we use $5000$ samples to generate one noise trajectory. Since \textit{dsp.ColoredNoise} gives noise trajectories with zero mean value, we additionally add a constant offset to every noise trajectory, to account for low frequency shifts. For every noise trajectory this offset is again obtained by a Gaussian sampling with mean $0$ and variance $A \log(T_{\rm exp}/T_{\rm  wf})$. Here $A$ is the amplitude of the power spectral density $S(f) = A/|f|$, $T_{\rm wf}$ is the duration of the actual readout process (here $10$ ns) and $T_{\rm exp} = N_r(T_{\rm wf}+T_{\rm reset})$ is the total experimental time. This time results from the sum of the actual measurement time plus the reset time $T_{\rm reset}$, multiplied by the number of repetitions $N_r$ necessary to obtain good measurement statistics. In our simulations we assume $T_{\rm reset} = 1$ ms, $N_r = 100$ and $A = (2\mu\Phi_0)^2$ for the small loop and $A = (10\mu\Phi_0)^2$ for the larger loop. These values are good upper bounds for realistic flux qubit experiments (e.g. \cite{PhysRevLett.99.187006}). 
\begin{figure}
\includegraphics[width=.4\textwidth]{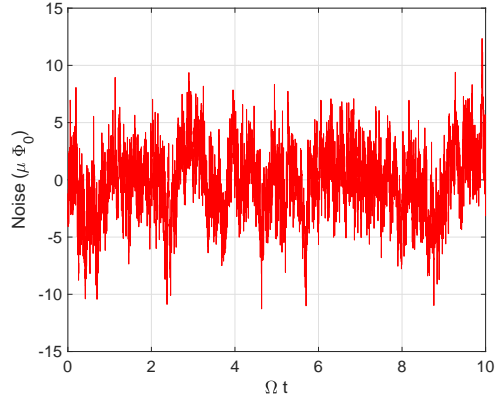}
\caption{Example of a $1/f$-noise signal generated by sampling Gaussian random processes and adding a variance to account for low frequency shifts.}
\label{noise1}
\end{figure}

In Fig. \ref{noise_cjj} we show the measurement fidelity for the 100 different $1/f$-noise generations. The results are shown for a qubit at (left) and away from the symmetry point (right). We see that in both cases the measurement fidelity is only changed in a very small amount ($\approx 10^{-3}$). This means that the additional ingredient of our scheme, i.e. the quantum probe, does not make the system more susceptible for flux noise.

Besides a direct change of the measurement results, flux noise induced in the cjj-SQUID could also lead to back action on the qubit itself. To prove that this also has no significant effect in our case, we study the time evolution of the qubit density matrix elements for one $1/f$-noise generation. As an input state we choose the persistent current state $\ket{\circlearrowleft}$ and the parameters are the same as in Fig. 3 of the main text. Fig. \ref{backaction_1f} shows the evolution of the density matrix element $\rho_{\circlearrowleft \circlearrowleft}(t)$ with and without additional $1/f$ flux noise in the two loops of the cjj-SQUID. We see that the QND fidelity of the qubit is only slightly affected by the noise, almost not visible in the figure. In Fig. \ref{backaction_1f} (right) we show an example of a $1/f$-noise signal generated by the algorithm and additionally the corresponding power spectral density to prove the 1/f behavior. 

All in all this proves that the additional circuit ingredient, i.e. the cjj-SQUID, does not make the measurement scheme more susceptible for typical noises appearing in flux qubit designs. 
\begin{figure}
\includegraphics[width=.4\textwidth]{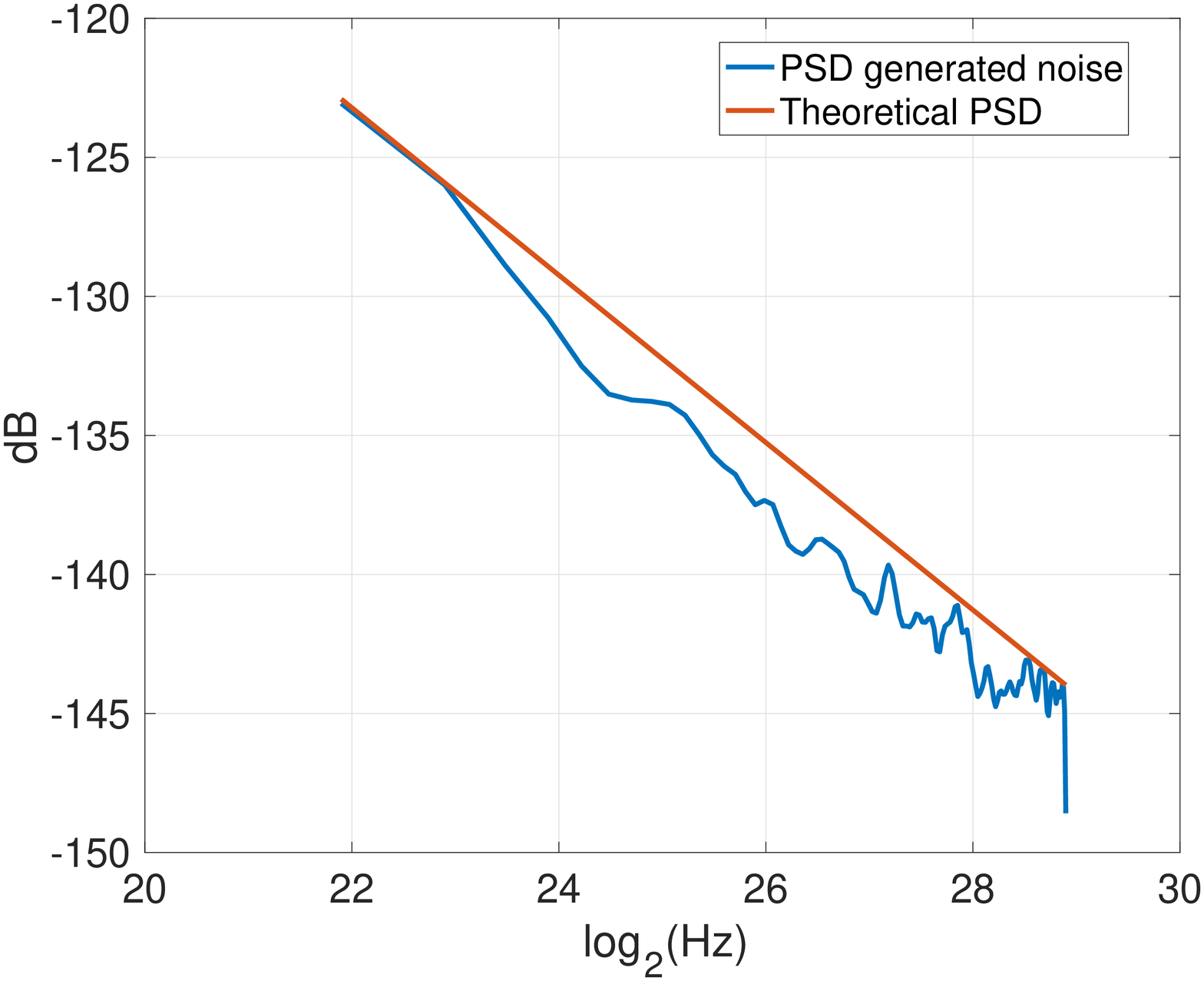}
\caption{Corresponding PSD to the 1/f noise signal shown in Fig. \ref{noise1}.}
\label{noise2}
\end{figure}

\end{appendix}

\bibliography{Bibliography.bib}

\begin{thebibliography}{51}%
\makeatletter
\providecommand \@ifxundefined [1]{%
 \@ifx{#1\undefined}
}%
\providecommand \@ifnum [1]{%
 \ifnum #1\expandafter \@firstoftwo
 \else \expandafter \@secondoftwo
 \fi
}%
\providecommand \@ifx [1]{%
 \ifx #1\expandafter \@firstoftwo
 \else \expandafter \@secondoftwo
 \fi
}%
\providecommand \natexlab [1]{#1}%
\providecommand \enquote  [1]{``#1''}%
\providecommand \bibnamefont  [1]{#1}%
\providecommand \bibfnamefont [1]{#1}%
\providecommand \citenamefont [1]{#1}%
\providecommand \href@noop [0]{\@secondoftwo}%
\providecommand \href [0]{\begingroup \@sanitize@url \@href}%
\providecommand \@href[1]{\@@startlink{#1}\@@href}%
\providecommand \@@href[1]{\endgroup#1\@@endlink}%
\providecommand \@sanitize@url [0]{\catcode `\\12\catcode `\$12\catcode
  `\&12\catcode `\#12\catcode `\^12\catcode `\_12\catcode `\%12\relax}%
\providecommand \@@startlink[1]{}%
\providecommand \@@endlink[0]{}%
\providecommand \url  [0]{\begingroup\@sanitize@url \@url }%
\providecommand \@url [1]{\endgroup\@href {#1}{\urlprefix }}%
\providecommand \urlprefix  [0]{URL }%
\providecommand \Eprint [0]{\href }%
\providecommand \doibase [0]{http://dx.doi.org/}%
\providecommand \selectlanguage [0]{\@gobble}%
\providecommand \bibinfo  [0]{\@secondoftwo}%
\providecommand \bibfield  [0]{\@secondoftwo}%
\providecommand \translation [1]{[#1]}%
\providecommand \BibitemOpen [0]{}%
\providecommand \bibitemStop [0]{}%
\providecommand \bibitemNoStop [0]{.\EOS\space}%
\providecommand \EOS [0]{\spacefactor3000\relax}%
\providecommand \BibitemShut  [1]{\csname bibitem#1\endcsname}%
\let\auto@bib@innerbib\@empty
\bibitem [{\citenamefont {Nielsen}\ and\ \citenamefont
  {Chuang}(2010)}]{NielsenChuan}%
  \BibitemOpen
  \bibfield  {author} {\bibinfo {author} {\bibfnamefont {M.~A.}\ \bibnamefont
  {Nielsen}}\ and\ \bibinfo {author} {\bibfnamefont {I.~L.}\ \bibnamefont
  {Chuang}},\ }\href@noop {} {\emph {\bibinfo {title} {Quantum Computation and
  Quantum Information}}}\ (\bibinfo  {publisher} {Cambridge University Press},\
  \bibinfo {year} {2010})\BibitemShut {NoStop}%
\bibitem [{\citenamefont {Braginsky}\ \emph {et~al.}(1980)\citenamefont
  {Braginsky}, \citenamefont {Vorontsov},\ and\ \citenamefont
  {Thorne}}]{braginsky1980quantum}%
  \BibitemOpen
  \bibfield  {author} {\bibinfo {author} {\bibfnamefont {V.~B.}\ \bibnamefont
  {Braginsky}}, \bibinfo {author} {\bibfnamefont {Y.~I.}\ \bibnamefont
  {Vorontsov}}, \ and\ \bibinfo {author} {\bibfnamefont {K.~S.}\ \bibnamefont
  {Thorne}},\ }\href@noop {} {\bibfield  {journal} {\bibinfo  {journal}
  {Science}\ }\textbf {\bibinfo {volume} {209}},\ \bibinfo {pages} {547}
  (\bibinfo {year} {1980})}\BibitemShut {NoStop}%
\bibitem [{\citenamefont {Braginsky}\ and\ \citenamefont
  {Khalili}(1996)}]{braginsky1996quantum}%
  \BibitemOpen
  \bibfield  {author} {\bibinfo {author} {\bibfnamefont {V.~B.}\ \bibnamefont
  {Braginsky}}\ and\ \bibinfo {author} {\bibfnamefont {F.~Y.}\ \bibnamefont
  {Khalili}},\ }\href@noop {} {\bibfield  {journal} {\bibinfo  {journal}
  {Reviews of Modern Physics}\ }\textbf {\bibinfo {volume} {68}},\ \bibinfo
  {pages} {1} (\bibinfo {year} {1996})}\BibitemShut {NoStop}%
\bibitem [{\citenamefont {Danilishin}\ and\ \citenamefont
  {Khalili}(2012)}]{danilishin2012quantum}%
  \BibitemOpen
  \bibfield  {author} {\bibinfo {author} {\bibfnamefont {S.~L.}\ \bibnamefont
  {Danilishin}}\ and\ \bibinfo {author} {\bibfnamefont {F.~Y.}\ \bibnamefont
  {Khalili}},\ }\href@noop {} {\bibfield  {journal} {\bibinfo  {journal}
  {Living Reviews in Relativity}\ }\textbf {\bibinfo {volume} {15}},\ \bibinfo
  {pages} {5} (\bibinfo {year} {2012})}\BibitemShut {NoStop}%
\bibitem [{\citenamefont {Fowler}\ \emph {et~al.}(2012)\citenamefont {Fowler},
  \citenamefont {Mariantoni}, \citenamefont {Martinis},\ and\ \citenamefont
  {Cleland}}]{fowler2012surface}%
  \BibitemOpen
  \bibfield  {author} {\bibinfo {author} {\bibfnamefont {A.~G.}\ \bibnamefont
  {Fowler}}, \bibinfo {author} {\bibfnamefont {M.}~\bibnamefont {Mariantoni}},
  \bibinfo {author} {\bibfnamefont {J.~M.}\ \bibnamefont {Martinis}}, \ and\
  \bibinfo {author} {\bibfnamefont {A.~N.}\ \bibnamefont {Cleland}},\
  }\href@noop {} {\bibfield  {journal} {\bibinfo  {journal} {Physical Review
  A}\ }\textbf {\bibinfo {volume} {86}},\ \bibinfo {pages} {032324} (\bibinfo
  {year} {2012})}\BibitemShut {NoStop}%
\bibitem [{\citenamefont {Ruskov}\ and\ \citenamefont
  {Korotkov}(2003)}]{ruskov2003entanglement}%
  \BibitemOpen
  \bibfield  {author} {\bibinfo {author} {\bibfnamefont {R.}~\bibnamefont
  {Ruskov}}\ and\ \bibinfo {author} {\bibfnamefont {A.~N.}\ \bibnamefont
  {Korotkov}},\ }\href@noop {} {\bibfield  {journal} {\bibinfo  {journal}
  {Physical Review B}\ }\textbf {\bibinfo {volume} {67}},\ \bibinfo {pages}
  {241305(R)} (\bibinfo {year} {2003})}\BibitemShut {NoStop}%
\bibitem [{\citenamefont {Mooij}\ \emph {et~al.}(1999)\citenamefont {Mooij},
  \citenamefont {Orlando}, \citenamefont {Levitov}, \citenamefont {Tian},
  \citenamefont {Van~der Wal},\ and\ \citenamefont
  {Lloyd}}]{mooij1999josephson}%
  \BibitemOpen
  \bibfield  {author} {\bibinfo {author} {\bibfnamefont {J.}~\bibnamefont
  {Mooij}}, \bibinfo {author} {\bibfnamefont {T.}~\bibnamefont {Orlando}},
  \bibinfo {author} {\bibfnamefont {L.}~\bibnamefont {Levitov}}, \bibinfo
  {author} {\bibfnamefont {L.}~\bibnamefont {Tian}}, \bibinfo {author}
  {\bibfnamefont {C.~H.}\ \bibnamefont {Van~der Wal}}, \ and\ \bibinfo {author}
  {\bibfnamefont {S.}~\bibnamefont {Lloyd}},\ }\href@noop {} {\bibfield
  {journal} {\bibinfo  {journal} {Science}\ }\textbf {\bibinfo {volume}
  {285}},\ \bibinfo {pages} {1036} (\bibinfo {year} {1999})}\BibitemShut
  {NoStop}%
\bibitem [{\citenamefont {Farhi}\ \emph {et~al.}(2001)\citenamefont {Farhi},
  \citenamefont {Goldstone}, \citenamefont {Gutmann}, \citenamefont {Lapan},
  \citenamefont {Lundgren},\ and\ \citenamefont {Preda}}]{farhi2001quantum}%
  \BibitemOpen
  \bibfield  {author} {\bibinfo {author} {\bibfnamefont {E.}~\bibnamefont
  {Farhi}}, \bibinfo {author} {\bibfnamefont {J.}~\bibnamefont {Goldstone}},
  \bibinfo {author} {\bibfnamefont {S.}~\bibnamefont {Gutmann}}, \bibinfo
  {author} {\bibfnamefont {J.}~\bibnamefont {Lapan}}, \bibinfo {author}
  {\bibfnamefont {A.}~\bibnamefont {Lundgren}}, \ and\ \bibinfo {author}
  {\bibfnamefont {D.}~\bibnamefont {Preda}},\ }\href@noop {} {\bibfield
  {journal} {\bibinfo  {journal} {Science}\ }\textbf {\bibinfo {volume}
  {292}},\ \bibinfo {pages} {472} (\bibinfo {year} {2001})}\BibitemShut
  {NoStop}%
\bibitem [{\citenamefont {Jansen}\ \emph {et~al.}(2007)\citenamefont {Jansen},
  \citenamefont {Ruskai},\ and\ \citenamefont {Seiler}}]{jansen2007bounds}%
  \BibitemOpen
  \bibfield  {author} {\bibinfo {author} {\bibfnamefont {S.}~\bibnamefont
  {Jansen}}, \bibinfo {author} {\bibfnamefont {M.-B.}\ \bibnamefont {Ruskai}},
  \ and\ \bibinfo {author} {\bibfnamefont {R.}~\bibnamefont {Seiler}},\
  }\href@noop {} {\bibfield  {journal} {\bibinfo  {journal} {Journal of
  Mathematical Physics}\ }\textbf {\bibinfo {volume} {48}},\ \bibinfo {pages}
  {102111} (\bibinfo {year} {2007})}\BibitemShut {NoStop}%
\bibitem [{\citenamefont {Aharonov}\ \emph {et~al.}(2008)\citenamefont
  {Aharonov}, \citenamefont {Van~Dam}, \citenamefont {Kempe}, \citenamefont
  {Landau}, \citenamefont {Lloyd},\ and\ \citenamefont
  {Regev}}]{aharonov2008adiabatic}%
  \BibitemOpen
  \bibfield  {author} {\bibinfo {author} {\bibfnamefont {D.}~\bibnamefont
  {Aharonov}}, \bibinfo {author} {\bibfnamefont {W.}~\bibnamefont {Van~Dam}},
  \bibinfo {author} {\bibfnamefont {J.}~\bibnamefont {Kempe}}, \bibinfo
  {author} {\bibfnamefont {Z.}~\bibnamefont {Landau}}, \bibinfo {author}
  {\bibfnamefont {S.}~\bibnamefont {Lloyd}}, \ and\ \bibinfo {author}
  {\bibfnamefont {O.}~\bibnamefont {Regev}},\ }\href@noop {} {\bibfield
  {journal} {\bibinfo  {journal} {SIAM review}\ }\textbf {\bibinfo {volume}
  {50}},\ \bibinfo {pages} {755} (\bibinfo {year} {2008})}\BibitemShut
  {NoStop}%
\bibitem [{\citenamefont {Bravyi}\ \emph {et~al.}(2008)\citenamefont {Bravyi},
  \citenamefont {DiVincenzo}, \citenamefont {Loss},\ and\ \citenamefont
  {Terhal}}]{bravyi2008quantum}%
  \BibitemOpen
  \bibfield  {author} {\bibinfo {author} {\bibfnamefont {S.}~\bibnamefont
  {Bravyi}}, \bibinfo {author} {\bibfnamefont {D.~P.}\ \bibnamefont
  {DiVincenzo}}, \bibinfo {author} {\bibfnamefont {D.}~\bibnamefont {Loss}}, \
  and\ \bibinfo {author} {\bibfnamefont {B.~M.}\ \bibnamefont {Terhal}},\
  }\href@noop {} {\bibfield  {journal} {\bibinfo  {journal} {Physical review
  letters}\ }\textbf {\bibinfo {volume} {101}},\ \bibinfo {pages} {070503}
  (\bibinfo {year} {2008})}\BibitemShut {NoStop}%
\bibitem [{\citenamefont {Amin}(2009)}]{amin2009consistency}%
  \BibitemOpen
  \bibfield  {author} {\bibinfo {author} {\bibfnamefont {M.}~\bibnamefont
  {Amin}},\ }\href@noop {} {\bibfield  {journal} {\bibinfo  {journal} {Physical
  review letters}\ }\textbf {\bibinfo {volume} {102}},\ \bibinfo {pages}
  {220401} (\bibinfo {year} {2009})}\BibitemShut {NoStop}%
\bibitem [{\citenamefont {Albash}\ and\ \citenamefont
  {Lidar}(2018)}]{RevModPhys.90.015002}%
  \BibitemOpen
  \bibfield  {author} {\bibinfo {author} {\bibfnamefont {T.}~\bibnamefont
  {Albash}}\ and\ \bibinfo {author} {\bibfnamefont {D.~A.}\ \bibnamefont
  {Lidar}},\ }\href@noop {} {\bibfield  {journal} {\bibinfo  {journal} {Rev.
  Mod. Phys.}\ }\textbf {\bibinfo {volume} {90}},\ \bibinfo {pages} {015002}
  (\bibinfo {year} {2018})}\BibitemShut {NoStop}%
\bibitem [{\citenamefont {Harris}\ \emph {et~al.}(2010)\citenamefont {Harris},
  \citenamefont {Johansson}, \citenamefont {Berkley}, \citenamefont {Johnson},
  \citenamefont {Lanting}, \citenamefont {Han}, \citenamefont {Bunyk},
  \citenamefont {Ladizinsky}, \citenamefont {Oh}, \citenamefont {Perminov},
  \citenamefont {Tolkacheva}, \citenamefont {Uchaikin}, \citenamefont
  {Chapple}, \citenamefont {Enderud}, \citenamefont {Rich}, \citenamefont
  {Thom}, \citenamefont {Wang}, \citenamefont {Wilson},\ and\ \citenamefont
  {Rose}}]{PhysRevB.81.134510}%
  \BibitemOpen
  \bibfield  {author} {\bibinfo {author} {\bibfnamefont {R.}~\bibnamefont
  {Harris}}, \bibinfo {author} {\bibfnamefont {J.}~\bibnamefont {Johansson}},
  \bibinfo {author} {\bibfnamefont {A.~J.}\ \bibnamefont {Berkley}}, \bibinfo
  {author} {\bibfnamefont {M.~W.}\ \bibnamefont {Johnson}}, \bibinfo {author}
  {\bibfnamefont {T.}~\bibnamefont {Lanting}}, \bibinfo {author} {\bibfnamefont
  {S.}~\bibnamefont {Han}}, \bibinfo {author} {\bibfnamefont {P.}~\bibnamefont
  {Bunyk}}, \bibinfo {author} {\bibfnamefont {E.}~\bibnamefont {Ladizinsky}},
  \bibinfo {author} {\bibfnamefont {T.}~\bibnamefont {Oh}}, \bibinfo {author}
  {\bibfnamefont {I.}~\bibnamefont {Perminov}}, \bibinfo {author}
  {\bibfnamefont {E.}~\bibnamefont {Tolkacheva}}, \bibinfo {author}
  {\bibfnamefont {S.}~\bibnamefont {Uchaikin}}, \bibinfo {author}
  {\bibfnamefont {E.~M.}\ \bibnamefont {Chapple}}, \bibinfo {author}
  {\bibfnamefont {C.}~\bibnamefont {Enderud}}, \bibinfo {author} {\bibfnamefont
  {C.}~\bibnamefont {Rich}}, \bibinfo {author} {\bibfnamefont {M.}~\bibnamefont
  {Thom}}, \bibinfo {author} {\bibfnamefont {J.}~\bibnamefont {Wang}}, \bibinfo
  {author} {\bibfnamefont {B.}~\bibnamefont {Wilson}}, \ and\ \bibinfo {author}
  {\bibfnamefont {G.}~\bibnamefont {Rose}},\ }\href {\doibase
  10.1103/PhysRevB.81.134510} {\bibfield  {journal} {\bibinfo  {journal} {Phys.
  Rev. B}\ }\textbf {\bibinfo {volume} {81}},\ \bibinfo {pages} {134510}
  (\bibinfo {year} {2010})}\BibitemShut {NoStop}%
\bibitem [{\citenamefont {Weber}\ \emph {et~al.}(2017)\citenamefont {Weber},
  \citenamefont {Samach}, \citenamefont {Hover}, \citenamefont {Gustavsson},
  \citenamefont {Kim}, \citenamefont {Melville}, \citenamefont {Rosenberg},
  \citenamefont {Sears}, \citenamefont {Yan}, \citenamefont {Yoder},
  \citenamefont {Oliver},\ and\ \citenamefont
  {Kerman}}]{PhysRevApplied.8.014004}%
  \BibitemOpen
  \bibfield  {author} {\bibinfo {author} {\bibfnamefont {S.~J.}\ \bibnamefont
  {Weber}}, \bibinfo {author} {\bibfnamefont {G.~O.}\ \bibnamefont {Samach}},
  \bibinfo {author} {\bibfnamefont {D.}~\bibnamefont {Hover}}, \bibinfo
  {author} {\bibfnamefont {S.}~\bibnamefont {Gustavsson}}, \bibinfo {author}
  {\bibfnamefont {D.~K.}\ \bibnamefont {Kim}}, \bibinfo {author} {\bibfnamefont
  {A.}~\bibnamefont {Melville}}, \bibinfo {author} {\bibfnamefont
  {D.}~\bibnamefont {Rosenberg}}, \bibinfo {author} {\bibfnamefont {A.~P.}\
  \bibnamefont {Sears}}, \bibinfo {author} {\bibfnamefont {F.}~\bibnamefont
  {Yan}}, \bibinfo {author} {\bibfnamefont {J.~L.}\ \bibnamefont {Yoder}},
  \bibinfo {author} {\bibfnamefont {W.~D.}\ \bibnamefont {Oliver}}, \ and\
  \bibinfo {author} {\bibfnamefont {A.~J.}\ \bibnamefont {Kerman}},\ }\href
  {\doibase 10.1103/PhysRevApplied.8.014004} {\bibfield  {journal} {\bibinfo
  {journal} {Phys. Rev. Applied}\ }\textbf {\bibinfo {volume} {8}},\ \bibinfo
  {pages} {014004} (\bibinfo {year} {2017})}\BibitemShut {NoStop}%
\bibitem [{\citenamefont {Lupa{\c{s}}cu}\ \emph {et~al.}(2005)\citenamefont
  {Lupa{\c{s}}cu}, \citenamefont {Harmans},\ and\ \citenamefont
  {Mooij}}]{lupacscu2005quantum}%
  \BibitemOpen
  \bibfield  {author} {\bibinfo {author} {\bibfnamefont {A.}~\bibnamefont
  {Lupa{\c{s}}cu}}, \bibinfo {author} {\bibfnamefont {C.}~\bibnamefont
  {Harmans}}, \ and\ \bibinfo {author} {\bibfnamefont {J.}~\bibnamefont
  {Mooij}},\ }\href@noop {} {\bibfield  {journal} {\bibinfo  {journal}
  {Physical Review B}\ }\textbf {\bibinfo {volume} {71}},\ \bibinfo {pages}
  {184506} (\bibinfo {year} {2005})}\BibitemShut {NoStop}%
\bibitem [{\citenamefont {Lupa{\c{s}}cu}\ \emph {et~al.}(2006)\citenamefont
  {Lupa{\c{s}}cu}, \citenamefont {Driessen}, \citenamefont {Roschier},
  \citenamefont {Harmans},\ and\ \citenamefont {Mooij}}]{lupacscu2006high}%
  \BibitemOpen
  \bibfield  {author} {\bibinfo {author} {\bibfnamefont {A.}~\bibnamefont
  {Lupa{\c{s}}cu}}, \bibinfo {author} {\bibfnamefont {E.~F.~C.}\ \bibnamefont
  {Driessen}}, \bibinfo {author} {\bibfnamefont {L.}~\bibnamefont {Roschier}},
  \bibinfo {author} {\bibfnamefont {C.~J. P.~M.}\ \bibnamefont {Harmans}}, \
  and\ \bibinfo {author} {\bibfnamefont {J.~E.}\ \bibnamefont {Mooij}},\
  }\href@noop {} {\bibfield  {journal} {\bibinfo  {journal} {Physical review
  letters}\ }\textbf {\bibinfo {volume} {96}},\ \bibinfo {pages} {127003}
  (\bibinfo {year} {2006})}\BibitemShut {NoStop}%
\bibitem [{\citenamefont {Boulant}\ \emph {et~al.}(2007)\citenamefont
  {Boulant}, \citenamefont {Ithier}, \citenamefont {Meeson}, \citenamefont
  {Nguyen}, \citenamefont {Vion}, \citenamefont {Esteve}, \citenamefont
  {Siddiqi}, \citenamefont {Vijay}, \citenamefont {Rigetti}, \citenamefont
  {Pierre} \emph {et~al.}}]{boulant2007quantum}%
  \BibitemOpen
  \bibfield  {author} {\bibinfo {author} {\bibfnamefont {N.}~\bibnamefont
  {Boulant}}, \bibinfo {author} {\bibfnamefont {G.}~\bibnamefont {Ithier}},
  \bibinfo {author} {\bibfnamefont {P.}~\bibnamefont {Meeson}}, \bibinfo
  {author} {\bibfnamefont {F.}~\bibnamefont {Nguyen}}, \bibinfo {author}
  {\bibfnamefont {D.}~\bibnamefont {Vion}}, \bibinfo {author} {\bibfnamefont
  {D.}~\bibnamefont {Esteve}}, \bibinfo {author} {\bibfnamefont
  {I.}~\bibnamefont {Siddiqi}}, \bibinfo {author} {\bibfnamefont
  {R.}~\bibnamefont {Vijay}}, \bibinfo {author} {\bibfnamefont
  {C.}~\bibnamefont {Rigetti}}, \bibinfo {author} {\bibfnamefont
  {F.}~\bibnamefont {Pierre}},  \emph {et~al.},\ }\href@noop {} {\bibfield
  {journal} {\bibinfo  {journal} {Physical Review B}\ }\textbf {\bibinfo
  {volume} {76}},\ \bibinfo {pages} {014525} (\bibinfo {year}
  {2007})}\BibitemShut {NoStop}%
\bibitem [{\citenamefont {Mallet}\ \emph {et~al.}(2009)\citenamefont {Mallet},
  \citenamefont {Ong}, \citenamefont {Palacios-Laloy}, \citenamefont {Nguyen},
  \citenamefont {Bertet}, \citenamefont {Vion},\ and\ \citenamefont
  {Esteve}}]{mallet2009single}%
  \BibitemOpen
  \bibfield  {author} {\bibinfo {author} {\bibfnamefont {F.}~\bibnamefont
  {Mallet}}, \bibinfo {author} {\bibfnamefont {F.~R.}\ \bibnamefont {Ong}},
  \bibinfo {author} {\bibfnamefont {A.}~\bibnamefont {Palacios-Laloy}},
  \bibinfo {author} {\bibfnamefont {F.}~\bibnamefont {Nguyen}}, \bibinfo
  {author} {\bibfnamefont {P.}~\bibnamefont {Bertet}}, \bibinfo {author}
  {\bibfnamefont {D.}~\bibnamefont {Vion}}, \ and\ \bibinfo {author}
  {\bibfnamefont {D.}~\bibnamefont {Esteve}},\ }\href@noop {} {\bibfield
  {journal} {\bibinfo  {journal} {Nature Physics}\ }\textbf {\bibinfo {volume}
  {5}},\ \bibinfo {pages} {791} (\bibinfo {year} {2009})}\BibitemShut {NoStop}%
\bibitem [{\citenamefont {Lupa\ifmmode~\mbox{\c{s}}\else \c{s}\fi{}cu}\ \emph
  {et~al.}(2004)\citenamefont {Lupa\ifmmode~\mbox{\c{s}}\else \c{s}\fi{}cu},
  \citenamefont {Verwijs}, \citenamefont {Schouten}, \citenamefont {Harmans},\
  and\ \citenamefont {Mooij}}]{PhysRevLett.93.177006}%
  \BibitemOpen
  \bibfield  {author} {\bibinfo {author} {\bibfnamefont {A.}~\bibnamefont
  {Lupa\ifmmode~\mbox{\c{s}}\else \c{s}\fi{}cu}}, \bibinfo {author}
  {\bibfnamefont {C.~J.~M.}\ \bibnamefont {Verwijs}}, \bibinfo {author}
  {\bibfnamefont {R.~N.}\ \bibnamefont {Schouten}}, \bibinfo {author}
  {\bibfnamefont {C.~J. P.~M.}\ \bibnamefont {Harmans}}, \ and\ \bibinfo
  {author} {\bibfnamefont {J.~E.}\ \bibnamefont {Mooij}},\ }\href {\doibase
  10.1103/PhysRevLett.93.177006} {\bibfield  {journal} {\bibinfo  {journal}
  {Phys. Rev. Lett.}\ }\textbf {\bibinfo {volume} {93}},\ \bibinfo {pages}
  {177006} (\bibinfo {year} {2004})}\BibitemShut {NoStop}%
\bibitem [{\citenamefont {Orgiazzi}\ \emph {et~al.}(2016)\citenamefont
  {Orgiazzi}, \citenamefont {Deng}, \citenamefont {Layden}, \citenamefont
  {Marchildon}, \citenamefont {Kitapli}, \citenamefont {Shen}, \citenamefont
  {Bal}, \citenamefont {Ong},\ and\ \citenamefont
  {Lupascu}}]{orgiazzi2016flux}%
  \BibitemOpen
  \bibfield  {author} {\bibinfo {author} {\bibfnamefont {J.-L.}\ \bibnamefont
  {Orgiazzi}}, \bibinfo {author} {\bibfnamefont {C.}~\bibnamefont {Deng}},
  \bibinfo {author} {\bibfnamefont {D.}~\bibnamefont {Layden}}, \bibinfo
  {author} {\bibfnamefont {R.}~\bibnamefont {Marchildon}}, \bibinfo {author}
  {\bibfnamefont {F.}~\bibnamefont {Kitapli}}, \bibinfo {author} {\bibfnamefont
  {F.}~\bibnamefont {Shen}}, \bibinfo {author} {\bibfnamefont {M.}~\bibnamefont
  {Bal}}, \bibinfo {author} {\bibfnamefont {F.~R.}\ \bibnamefont {Ong}}, \ and\
  \bibinfo {author} {\bibfnamefont {A.}~\bibnamefont {Lupascu}},\ }\href@noop
  {} {\bibfield  {journal} {\bibinfo  {journal} {Physical Review B}\ }\textbf
  {\bibinfo {volume} {93}},\ \bibinfo {pages} {104518} (\bibinfo {year}
  {2016})}\BibitemShut {NoStop}%
\bibitem [{\citenamefont {Inomata}\ \emph {et~al.}(2012)\citenamefont
  {Inomata}, \citenamefont {Yamamoto}, \citenamefont {Billangeon},
  \citenamefont {Nakamura},\ and\ \citenamefont {Tsai}}]{inomata2012large}%
  \BibitemOpen
  \bibfield  {author} {\bibinfo {author} {\bibfnamefont {K.}~\bibnamefont
  {Inomata}}, \bibinfo {author} {\bibfnamefont {T.}~\bibnamefont {Yamamoto}},
  \bibinfo {author} {\bibfnamefont {P.-M.}\ \bibnamefont {Billangeon}},
  \bibinfo {author} {\bibfnamefont {Y.}~\bibnamefont {Nakamura}}, \ and\
  \bibinfo {author} {\bibfnamefont {J.~S.}\ \bibnamefont {Tsai}},\ }\href@noop
  {} {\bibfield  {journal} {\bibinfo  {journal} {Physical Review B}\ }\textbf
  {\bibinfo {volume} {86}},\ \bibinfo {pages} {140508} (\bibinfo {year}
  {2012})}\BibitemShut {NoStop}%
\bibitem [{\citenamefont {Rifkin}\ and\ \citenamefont
  {Deaver~Jr}(1976)}]{rifkin1976current}%
  \BibitemOpen
  \bibfield  {author} {\bibinfo {author} {\bibfnamefont {R.}~\bibnamefont
  {Rifkin}}\ and\ \bibinfo {author} {\bibfnamefont {B.~S.}\ \bibnamefont
  {Deaver~Jr}},\ }\href@noop {} {\bibfield  {journal} {\bibinfo  {journal}
  {Physical Review B}\ }\textbf {\bibinfo {volume} {13}},\ \bibinfo {pages}
  {3894} (\bibinfo {year} {1976})}\BibitemShut {NoStop}%
\bibitem [{\citenamefont {Greenberg}\ \emph {et~al.}(2002)\citenamefont
  {Greenberg}, \citenamefont {Izmalkov}, \citenamefont {Grajcar}, \citenamefont
  {Il’ichev}, \citenamefont {Krech}, \citenamefont {Meyer}, \citenamefont
  {Amin},\ and\ \citenamefont {van~den Brink}}]{greenberg2002low}%
  \BibitemOpen
  \bibfield  {author} {\bibinfo {author} {\bibfnamefont {Y.~S.}\ \bibnamefont
  {Greenberg}}, \bibinfo {author} {\bibfnamefont {A.}~\bibnamefont {Izmalkov}},
  \bibinfo {author} {\bibfnamefont {M.}~\bibnamefont {Grajcar}}, \bibinfo
  {author} {\bibfnamefont {E.}~\bibnamefont {Il’ichev}}, \bibinfo {author}
  {\bibfnamefont {W.}~\bibnamefont {Krech}}, \bibinfo {author} {\bibfnamefont
  {H.-G.}\ \bibnamefont {Meyer}}, \bibinfo {author} {\bibfnamefont {M.~H.~S.}\
  \bibnamefont {Amin}}, \ and\ \bibinfo {author} {\bibfnamefont {A.~M.}\
  \bibnamefont {van~den Brink}},\ }\href@noop {} {\bibfield  {journal}
  {\bibinfo  {journal} {Physical Review B}\ }\textbf {\bibinfo {volume} {66}},\
  \bibinfo {pages} {214525} (\bibinfo {year} {2002})}\BibitemShut {NoStop}%
\bibitem [{\citenamefont {Wang}\ \emph {et~al.}(2011)\citenamefont {Wang},
  \citenamefont {Zhu},\ and\ \citenamefont {Bruder}}]{wang2011ideal}%
  \BibitemOpen
  \bibfield  {author} {\bibinfo {author} {\bibfnamefont {Y.-D.}\ \bibnamefont
  {Wang}}, \bibinfo {author} {\bibfnamefont {X.}~\bibnamefont {Zhu}}, \ and\
  \bibinfo {author} {\bibfnamefont {C.}~\bibnamefont {Bruder}},\ }\href@noop {}
  {\bibfield  {journal} {\bibinfo  {journal} {Physical Review B}\ }\textbf
  {\bibinfo {volume} {83}},\ \bibinfo {pages} {134504} (\bibinfo {year}
  {2011})}\BibitemShut {NoStop}%
\bibitem [{\citenamefont {Harris}\ \emph {et~al.}(2018)\citenamefont {Harris},
  \citenamefont {Sato}, \citenamefont {Berkley}, \citenamefont {Reis},
  \citenamefont {Altomare}, \citenamefont {Amin}, \citenamefont {Boothby},
  \citenamefont {Bunyk}, \citenamefont {Deng}, \citenamefont {Enderud} \emph
  {et~al.}}]{harris2018phase}%
  \BibitemOpen
  \bibfield  {author} {\bibinfo {author} {\bibfnamefont {R.}~\bibnamefont
  {Harris}}, \bibinfo {author} {\bibfnamefont {Y.}~\bibnamefont {Sato}},
  \bibinfo {author} {\bibfnamefont {A.}~\bibnamefont {Berkley}}, \bibinfo
  {author} {\bibfnamefont {M.}~\bibnamefont {Reis}}, \bibinfo {author}
  {\bibfnamefont {F.}~\bibnamefont {Altomare}}, \bibinfo {author}
  {\bibfnamefont {M.}~\bibnamefont {Amin}}, \bibinfo {author} {\bibfnamefont
  {K.}~\bibnamefont {Boothby}}, \bibinfo {author} {\bibfnamefont
  {P.}~\bibnamefont {Bunyk}}, \bibinfo {author} {\bibfnamefont
  {C.}~\bibnamefont {Deng}}, \bibinfo {author} {\bibfnamefont {C.}~\bibnamefont
  {Enderud}},  \emph {et~al.},\ }\href@noop {} {\bibfield  {journal} {\bibinfo
  {journal} {Science}\ }\textbf {\bibinfo {volume} {361}},\ \bibinfo {pages}
  {162} (\bibinfo {year} {2018})}\BibitemShut {NoStop}%
\bibitem [{\citenamefont {King}\ \emph {et~al.}(2018)\citenamefont {King},
  \citenamefont {Carrasquilla}, \citenamefont {Raymond}, \citenamefont
  {Ozfidan}, \citenamefont {Andriyash}, \citenamefont {Berkley}, \citenamefont
  {Reis}, \citenamefont {Lanting}, \citenamefont {Harris}, \citenamefont
  {Altomare} \emph {et~al.}}]{king2018observation}%
  \BibitemOpen
  \bibfield  {author} {\bibinfo {author} {\bibfnamefont {A.~D.}\ \bibnamefont
  {King}}, \bibinfo {author} {\bibfnamefont {J.}~\bibnamefont {Carrasquilla}},
  \bibinfo {author} {\bibfnamefont {J.}~\bibnamefont {Raymond}}, \bibinfo
  {author} {\bibfnamefont {I.}~\bibnamefont {Ozfidan}}, \bibinfo {author}
  {\bibfnamefont {E.}~\bibnamefont {Andriyash}}, \bibinfo {author}
  {\bibfnamefont {A.}~\bibnamefont {Berkley}}, \bibinfo {author} {\bibfnamefont
  {M.}~\bibnamefont {Reis}}, \bibinfo {author} {\bibfnamefont {T.}~\bibnamefont
  {Lanting}}, \bibinfo {author} {\bibfnamefont {R.}~\bibnamefont {Harris}},
  \bibinfo {author} {\bibfnamefont {F.}~\bibnamefont {Altomare}},  \emph
  {et~al.},\ }\href@noop {} {\bibfield  {journal} {\bibinfo  {journal}
  {Nature}\ }\textbf {\bibinfo {volume} {560}},\ \bibinfo {pages} {456}
  (\bibinfo {year} {2018})}\BibitemShut {NoStop}%
\bibitem [{\citenamefont {Fujii}(2018)}]{fujii2018quantum}%
  \BibitemOpen
  \bibfield  {author} {\bibinfo {author} {\bibfnamefont {K.}~\bibnamefont
  {Fujii}},\ }\href@noop {} {\bibfield  {journal} {\bibinfo  {journal} {arXiv
  preprint arXiv:1803.09954}\ } (\bibinfo {year} {2018})}\BibitemShut {NoStop}%
\bibitem [{\citenamefont {Lvovsky}\ and\ \citenamefont
  {Raymer}(2009)}]{lvovsky2009continuous}%
  \BibitemOpen
  \bibfield  {author} {\bibinfo {author} {\bibfnamefont {A.~I.}\ \bibnamefont
  {Lvovsky}}\ and\ \bibinfo {author} {\bibfnamefont {M.~G.}\ \bibnamefont
  {Raymer}},\ }\href@noop {} {\bibfield  {journal} {\bibinfo  {journal}
  {Reviews of Modern Physics}\ }\textbf {\bibinfo {volume} {81}},\ \bibinfo
  {pages} {299} (\bibinfo {year} {2009})}\BibitemShut {NoStop}%
\bibitem [{\citenamefont {Filipp}\ \emph {et~al.}(2009)\citenamefont {Filipp},
  \citenamefont {Maurer}, \citenamefont {Leek}, \citenamefont {Baur},
  \citenamefont {Bianchetti}, \citenamefont {Fink}, \citenamefont {G{\"o}ppl},
  \citenamefont {Steffen}, \citenamefont {Gambetta}, \citenamefont {Blais}
  \emph {et~al.}}]{filipp2009two}%
  \BibitemOpen
  \bibfield  {author} {\bibinfo {author} {\bibfnamefont {S.}~\bibnamefont
  {Filipp}}, \bibinfo {author} {\bibfnamefont {P.}~\bibnamefont {Maurer}},
  \bibinfo {author} {\bibfnamefont {P.~J.}\ \bibnamefont {Leek}}, \bibinfo
  {author} {\bibfnamefont {M.}~\bibnamefont {Baur}}, \bibinfo {author}
  {\bibfnamefont {R.}~\bibnamefont {Bianchetti}}, \bibinfo {author}
  {\bibfnamefont {J.~M.}\ \bibnamefont {Fink}}, \bibinfo {author}
  {\bibfnamefont {M.}~\bibnamefont {G{\"o}ppl}}, \bibinfo {author}
  {\bibfnamefont {L.}~\bibnamefont {Steffen}}, \bibinfo {author} {\bibfnamefont
  {J.~M.}\ \bibnamefont {Gambetta}}, \bibinfo {author} {\bibfnamefont
  {A.}~\bibnamefont {Blais}},  \emph {et~al.},\ }\href@noop {} {\bibfield
  {journal} {\bibinfo  {journal} {Physical review letters}\ }\textbf {\bibinfo
  {volume} {102}},\ \bibinfo {pages} {200402} (\bibinfo {year}
  {2009})}\BibitemShut {NoStop}%
\bibitem [{\citenamefont {Harris}\ \emph {et~al.}(2009)\citenamefont {Harris},
  \citenamefont {Lanting}, \citenamefont {Berkley}, \citenamefont {Johansson},
  \citenamefont {Johnson}, \citenamefont {Bunyk}, \citenamefont {Ladizinsky},
  \citenamefont {Ladizinsky}, \citenamefont {Oh},\ and\ \citenamefont
  {Han}}]{harris2009compound}%
  \BibitemOpen
  \bibfield  {author} {\bibinfo {author} {\bibfnamefont {R.}~\bibnamefont
  {Harris}}, \bibinfo {author} {\bibfnamefont {T.}~\bibnamefont {Lanting}},
  \bibinfo {author} {\bibfnamefont {A.~J.}\ \bibnamefont {Berkley}}, \bibinfo
  {author} {\bibfnamefont {J.}~\bibnamefont {Johansson}}, \bibinfo {author}
  {\bibfnamefont {M.~W.}\ \bibnamefont {Johnson}}, \bibinfo {author}
  {\bibfnamefont {P.}~\bibnamefont {Bunyk}}, \bibinfo {author} {\bibfnamefont
  {E.}~\bibnamefont {Ladizinsky}}, \bibinfo {author} {\bibfnamefont
  {N.}~\bibnamefont {Ladizinsky}}, \bibinfo {author} {\bibfnamefont
  {T.}~\bibnamefont {Oh}}, \ and\ \bibinfo {author} {\bibfnamefont
  {S.}~\bibnamefont {Han}},\ }\href@noop {} {\bibfield  {journal} {\bibinfo
  {journal} {Physical Review B}\ }\textbf {\bibinfo {volume} {80}},\ \bibinfo
  {pages} {052506} (\bibinfo {year} {2009})}\BibitemShut {NoStop}%
\bibitem [{\citenamefont {Poletto}\ \emph {et~al.}(2009)\citenamefont
  {Poletto}, \citenamefont {Chiarello}, \citenamefont {Castellano},
  \citenamefont {Lisenfeld}, \citenamefont {Lukashenko}, \citenamefont
  {Carelli},\ and\ \citenamefont {Ustinov}}]{poletto2009tunable}%
  \BibitemOpen
  \bibfield  {author} {\bibinfo {author} {\bibfnamefont {S.}~\bibnamefont
  {Poletto}}, \bibinfo {author} {\bibfnamefont {F.}~\bibnamefont {Chiarello}},
  \bibinfo {author} {\bibfnamefont {M.}~\bibnamefont {Castellano}}, \bibinfo
  {author} {\bibfnamefont {J.}~\bibnamefont {Lisenfeld}}, \bibinfo {author}
  {\bibfnamefont {A.}~\bibnamefont {Lukashenko}}, \bibinfo {author}
  {\bibfnamefont {P.}~\bibnamefont {Carelli}}, \ and\ \bibinfo {author}
  {\bibfnamefont {A.}~\bibnamefont {Ustinov}},\ }\href@noop {} {\bibfield
  {journal} {\bibinfo  {journal} {Physica Scripta}\ }\textbf {\bibinfo {volume}
  {2009}},\ \bibinfo {pages} {014011} (\bibinfo {year} {2009})}\BibitemShut
  {NoStop}%
\bibitem [{\citenamefont {Castellano}\ \emph {et~al.}(2010)\citenamefont
  {Castellano}, \citenamefont {Chiarello}, \citenamefont {Carelli},
  \citenamefont {Cosmelli}, \citenamefont {Mattioli},\ and\ \citenamefont
  {Torrioli}}]{castellano2010deep}%
  \BibitemOpen
  \bibfield  {author} {\bibinfo {author} {\bibfnamefont {M.}~\bibnamefont
  {Castellano}}, \bibinfo {author} {\bibfnamefont {F.}~\bibnamefont
  {Chiarello}}, \bibinfo {author} {\bibfnamefont {P.}~\bibnamefont {Carelli}},
  \bibinfo {author} {\bibfnamefont {C.}~\bibnamefont {Cosmelli}}, \bibinfo
  {author} {\bibfnamefont {F.}~\bibnamefont {Mattioli}}, \ and\ \bibinfo
  {author} {\bibfnamefont {G.}~\bibnamefont {Torrioli}},\ }\href@noop {}
  {\bibfield  {journal} {\bibinfo  {journal} {New Journal of Physics}\ }\textbf
  {\bibinfo {volume} {12}},\ \bibinfo {pages} {043047} (\bibinfo {year}
  {2010})}\BibitemShut {NoStop}%
\bibitem [{\citenamefont {Kockum}\ \emph {et~al.}(2019)\citenamefont {Kockum},
  \citenamefont {Miranowicz}, \citenamefont {De~Liberato}, \citenamefont
  {Savasta},\ and\ \citenamefont {Nori}}]{kockum2019ultrastrong}%
  \BibitemOpen
  \bibfield  {author} {\bibinfo {author} {\bibfnamefont {A.~F.}\ \bibnamefont
  {Kockum}}, \bibinfo {author} {\bibfnamefont {A.}~\bibnamefont {Miranowicz}},
  \bibinfo {author} {\bibfnamefont {S.}~\bibnamefont {De~Liberato}}, \bibinfo
  {author} {\bibfnamefont {S.}~\bibnamefont {Savasta}}, \ and\ \bibinfo
  {author} {\bibfnamefont {F.}~\bibnamefont {Nori}},\ }\href@noop {} {\bibfield
   {journal} {\bibinfo  {journal} {Nature Reviews Physics}\ }\textbf {\bibinfo
  {volume} {1}},\ \bibinfo {pages} {19} (\bibinfo {year} {2019})}\BibitemShut
  {NoStop}%
\bibitem [{\citenamefont {Wilhelm}(2003)}]{wilhelm2003asymptotic}%
  \BibitemOpen
  \bibfield  {author} {\bibinfo {author} {\bibfnamefont {F.~K.}\ \bibnamefont
  {Wilhelm}},\ }\href@noop {} {\bibfield  {journal} {\bibinfo  {journal}
  {Physical Review B}\ }\textbf {\bibinfo {volume} {68}},\ \bibinfo {pages}
  {060503(R)} (\bibinfo {year} {2003})}\BibitemShut {NoStop}%
\bibitem [{\citenamefont {Kato}(1950)}]{Kato1950}%
  \BibitemOpen
  \bibfield  {author} {\bibinfo {author} {\bibfnamefont {T.}~\bibnamefont
  {Kato}},\ }\href@noop {} {\bibfield  {journal} {\bibinfo  {journal} {Journal
  of the Physical Society of Japan}\ }\textbf {\bibinfo {volume} {5}},\
  \bibinfo {pages} {435} (\bibinfo {year} {1950})}\BibitemShut {NoStop}%
\bibitem [{\citenamefont {Braginsky}\ and\ \citenamefont
  {Khalili}(1992)}]{braginsky1992measurement}%
  \BibitemOpen
  \bibfield  {author} {\bibinfo {author} {\bibfnamefont {V.~B.}\ \bibnamefont
  {Braginsky}}\ and\ \bibinfo {author} {\bibfnamefont {F.~Y.}\ \bibnamefont
  {Khalili}},\ }\href@noop {} {\emph {\bibinfo {title} {Quantum measurement}}}\
  (\bibinfo  {publisher} {Cambridge University Press},\ \bibinfo {year}
  {1992})\BibitemShut {NoStop}%
\bibitem [{\citenamefont {Clerk}\ \emph {et~al.}(2003)\citenamefont {Clerk},
  \citenamefont {Girvin},\ and\ \citenamefont {Stone}}]{clerk2003quantum}%
  \BibitemOpen
  \bibfield  {author} {\bibinfo {author} {\bibfnamefont {A.}~\bibnamefont
  {Clerk}}, \bibinfo {author} {\bibfnamefont {S.}~\bibnamefont {Girvin}}, \
  and\ \bibinfo {author} {\bibfnamefont {A.~D.}\ \bibnamefont {Stone}},\
  }\href@noop {} {\bibfield  {journal} {\bibinfo  {journal} {Physical Review
  B}\ }\textbf {\bibinfo {volume} {67}},\ \bibinfo {pages} {165324} (\bibinfo
  {year} {2003})}\BibitemShut {NoStop}%
\bibitem [{\citenamefont {Allman}\ \emph {et~al.}(2010)\citenamefont {Allman},
  \citenamefont {Altomare}, \citenamefont {Whittaker}, \citenamefont {Cicak},
  \citenamefont {Li}, \citenamefont {Sirois}, \citenamefont {Strong},
  \citenamefont {Teufel},\ and\ \citenamefont
  {Simmonds}}]{PhysRevLett.104.177004}%
  \BibitemOpen
  \bibfield  {author} {\bibinfo {author} {\bibfnamefont {M.~S.}\ \bibnamefont
  {Allman}}, \bibinfo {author} {\bibfnamefont {F.}~\bibnamefont {Altomare}},
  \bibinfo {author} {\bibfnamefont {J.~D.}\ \bibnamefont {Whittaker}}, \bibinfo
  {author} {\bibfnamefont {K.}~\bibnamefont {Cicak}}, \bibinfo {author}
  {\bibfnamefont {D.}~\bibnamefont {Li}}, \bibinfo {author} {\bibfnamefont
  {A.}~\bibnamefont {Sirois}}, \bibinfo {author} {\bibfnamefont
  {J.}~\bibnamefont {Strong}}, \bibinfo {author} {\bibfnamefont {J.~D.}\
  \bibnamefont {Teufel}}, \ and\ \bibinfo {author} {\bibfnamefont {R.~W.}\
  \bibnamefont {Simmonds}},\ }\href@noop {} {\bibfield  {journal} {\bibinfo
  {journal} {Phys. Rev. Lett.}\ }\textbf {\bibinfo {volume} {104}},\ \bibinfo
  {pages} {177004} (\bibinfo {year} {2010})}\BibitemShut {NoStop}%
\bibitem [{\citenamefont {Quintana}\ \emph {et~al.}(2017)\citenamefont
  {Quintana}, \citenamefont {Chen}, \citenamefont {Sank}, \citenamefont
  {Petukhov}, \citenamefont {White}, \citenamefont {Kafri}, \citenamefont
  {Chiaro}, \citenamefont {Megrant}, \citenamefont {Barends}, \citenamefont
  {Campbell}, \citenamefont {Chen}, \citenamefont {Dunsworth}, \citenamefont
  {Fowler}, \citenamefont {Graff}, \citenamefont {Jeffrey}, \citenamefont
  {Kelly}, \citenamefont {Lucero}, \citenamefont {Mutus}, \citenamefont
  {Neeley}, \citenamefont {Neill}, \citenamefont {O'Malley}, \citenamefont
  {Roushan}, \citenamefont {Shabani}, \citenamefont {Smelyanskiy},
  \citenamefont {Vainsencher}, \citenamefont {Wenner}, \citenamefont {Neven},\
  and\ \citenamefont {Martinis}}]{PhysRevLett.118.057702}%
  \BibitemOpen
  \bibfield  {author} {\bibinfo {author} {\bibfnamefont {C.~M.}\ \bibnamefont
  {Quintana}}, \bibinfo {author} {\bibfnamefont {Y.}~\bibnamefont {Chen}},
  \bibinfo {author} {\bibfnamefont {D.}~\bibnamefont {Sank}}, \bibinfo {author}
  {\bibfnamefont {A.~G.}\ \bibnamefont {Petukhov}}, \bibinfo {author}
  {\bibfnamefont {T.~C.}\ \bibnamefont {White}}, \bibinfo {author}
  {\bibfnamefont {D.}~\bibnamefont {Kafri}}, \bibinfo {author} {\bibfnamefont
  {B.}~\bibnamefont {Chiaro}}, \bibinfo {author} {\bibfnamefont
  {A.}~\bibnamefont {Megrant}}, \bibinfo {author} {\bibfnamefont
  {R.}~\bibnamefont {Barends}}, \bibinfo {author} {\bibfnamefont
  {B.}~\bibnamefont {Campbell}}, \bibinfo {author} {\bibfnamefont
  {Z.}~\bibnamefont {Chen}}, \bibinfo {author} {\bibfnamefont {A.}~\bibnamefont
  {Dunsworth}}, \bibinfo {author} {\bibfnamefont {A.~G.}\ \bibnamefont
  {Fowler}}, \bibinfo {author} {\bibfnamefont {R.}~\bibnamefont {Graff}},
  \bibinfo {author} {\bibfnamefont {E.}~\bibnamefont {Jeffrey}}, \bibinfo
  {author} {\bibfnamefont {J.}~\bibnamefont {Kelly}}, \bibinfo {author}
  {\bibfnamefont {E.}~\bibnamefont {Lucero}}, \bibinfo {author} {\bibfnamefont
  {J.~Y.}\ \bibnamefont {Mutus}}, \bibinfo {author} {\bibfnamefont
  {M.}~\bibnamefont {Neeley}}, \bibinfo {author} {\bibfnamefont
  {C.}~\bibnamefont {Neill}}, \bibinfo {author} {\bibfnamefont {P.~J.~J.}\
  \bibnamefont {O'Malley}}, \bibinfo {author} {\bibfnamefont {P.}~\bibnamefont
  {Roushan}}, \bibinfo {author} {\bibfnamefont {A.}~\bibnamefont {Shabani}},
  \bibinfo {author} {\bibfnamefont {V.~N.}\ \bibnamefont {Smelyanskiy}},
  \bibinfo {author} {\bibfnamefont {A.}~\bibnamefont {Vainsencher}}, \bibinfo
  {author} {\bibfnamefont {J.}~\bibnamefont {Wenner}}, \bibinfo {author}
  {\bibfnamefont {H.}~\bibnamefont {Neven}}, \ and\ \bibinfo {author}
  {\bibfnamefont {J.~M.}\ \bibnamefont {Martinis}},\ }\href@noop {} {\bibfield
  {journal} {\bibinfo  {journal} {Phys. Rev. Lett.}\ }\textbf {\bibinfo
  {volume} {118}},\ \bibinfo {pages} {057702} (\bibinfo {year}
  {2017})}\BibitemShut {NoStop}%
\bibitem [{\citenamefont {Neill}\ \emph {et~al.}(2016)\citenamefont {Neill},
  \citenamefont {Roushan}, \citenamefont {Fang}, \citenamefont {Chen},
  \citenamefont {Kolodrubetz}, \citenamefont {Chen}, \citenamefont {Megrant},
  \citenamefont {Barends}, \citenamefont {Campbell}, \citenamefont {Chiaro}
  \emph {et~al.}}]{neill2016ergodic}%
  \BibitemOpen
  \bibfield  {author} {\bibinfo {author} {\bibfnamefont {C.}~\bibnamefont
  {Neill}}, \bibinfo {author} {\bibfnamefont {P.}~\bibnamefont {Roushan}},
  \bibinfo {author} {\bibfnamefont {M.}~\bibnamefont {Fang}}, \bibinfo {author}
  {\bibfnamefont {Y.}~\bibnamefont {Chen}}, \bibinfo {author} {\bibfnamefont
  {M.}~\bibnamefont {Kolodrubetz}}, \bibinfo {author} {\bibfnamefont
  {Z.}~\bibnamefont {Chen}}, \bibinfo {author} {\bibfnamefont {A.}~\bibnamefont
  {Megrant}}, \bibinfo {author} {\bibfnamefont {R.}~\bibnamefont {Barends}},
  \bibinfo {author} {\bibfnamefont {B.}~\bibnamefont {Campbell}}, \bibinfo
  {author} {\bibfnamefont {B.}~\bibnamefont {Chiaro}},  \emph {et~al.},\
  }\href@noop {} {\bibfield  {journal} {\bibinfo  {journal} {Nature Physics}\
  }\textbf {\bibinfo {volume} {12}},\ \bibinfo {pages} {1037} (\bibinfo {year}
  {2016})}\BibitemShut {NoStop}%
\bibitem [{\citenamefont {Niemczyk}\ \emph {et~al.}(2010)\citenamefont
  {Niemczyk}, \citenamefont {Deppe}, \citenamefont {Huebl}, \citenamefont
  {Menzel}, \citenamefont {Hocke}, \citenamefont {Schwarz}, \citenamefont
  {Garcia-Ripoll}, \citenamefont {Zueco}, \citenamefont {H{\"u}mmer},
  \citenamefont {Solano} \emph {et~al.}}]{niemczyk2010circuit}%
  \BibitemOpen
  \bibfield  {author} {\bibinfo {author} {\bibfnamefont {T.}~\bibnamefont
  {Niemczyk}}, \bibinfo {author} {\bibfnamefont {F.}~\bibnamefont {Deppe}},
  \bibinfo {author} {\bibfnamefont {H.}~\bibnamefont {Huebl}}, \bibinfo
  {author} {\bibfnamefont {E.}~\bibnamefont {Menzel}}, \bibinfo {author}
  {\bibfnamefont {F.}~\bibnamefont {Hocke}}, \bibinfo {author} {\bibfnamefont
  {M.}~\bibnamefont {Schwarz}}, \bibinfo {author} {\bibfnamefont
  {J.}~\bibnamefont {Garcia-Ripoll}}, \bibinfo {author} {\bibfnamefont
  {D.}~\bibnamefont {Zueco}}, \bibinfo {author} {\bibfnamefont
  {T.}~\bibnamefont {H{\"u}mmer}}, \bibinfo {author} {\bibfnamefont
  {E.}~\bibnamefont {Solano}},  \emph {et~al.},\ }\href@noop {} {\bibfield
  {journal} {\bibinfo  {journal} {Nature Physics}\ }\textbf {\bibinfo {volume}
  {6}},\ \bibinfo {pages} {772} (\bibinfo {year} {2010})}\BibitemShut {NoStop}%
\bibitem [{\citenamefont {Baust}\ \emph {et~al.}(2016)\citenamefont {Baust},
  \citenamefont {Hoffmann}, \citenamefont {Haeberlein}, \citenamefont
  {Schwarz}, \citenamefont {Eder}, \citenamefont {Goetz}, \citenamefont
  {Wulschner}, \citenamefont {Xie}, \citenamefont {Zhong}, \citenamefont
  {Quijandr{\'\i}a} \emph {et~al.}}]{baust2016ultrastrong}%
  \BibitemOpen
  \bibfield  {author} {\bibinfo {author} {\bibfnamefont {A.}~\bibnamefont
  {Baust}}, \bibinfo {author} {\bibfnamefont {E.}~\bibnamefont {Hoffmann}},
  \bibinfo {author} {\bibfnamefont {M.}~\bibnamefont {Haeberlein}}, \bibinfo
  {author} {\bibfnamefont {M.~J.}\ \bibnamefont {Schwarz}}, \bibinfo {author}
  {\bibfnamefont {P.}~\bibnamefont {Eder}}, \bibinfo {author} {\bibfnamefont
  {J.}~\bibnamefont {Goetz}}, \bibinfo {author} {\bibfnamefont
  {F.}~\bibnamefont {Wulschner}}, \bibinfo {author} {\bibfnamefont
  {E.}~\bibnamefont {Xie}}, \bibinfo {author} {\bibfnamefont {L.}~\bibnamefont
  {Zhong}}, \bibinfo {author} {\bibfnamefont {F.}~\bibnamefont
  {Quijandr{\'\i}a}},  \emph {et~al.},\ }\href@noop {} {\bibfield  {journal}
  {\bibinfo  {journal} {Physical Review B}\ }\textbf {\bibinfo {volume} {93}},\
  \bibinfo {pages} {214501} (\bibinfo {year} {2016})}\BibitemShut {NoStop}%
\bibitem [{\citenamefont {Forn-D{\'\i}az}\ \emph {et~al.}(2017)\citenamefont
  {Forn-D{\'\i}az}, \citenamefont {Garc{\'\i}a-Ripoll}, \citenamefont
  {Peropadre}, \citenamefont {Orgiazzi}, \citenamefont {Yurtalan},
  \citenamefont {Belyansky}, \citenamefont {Wilson},\ and\ \citenamefont
  {Lupascu}}]{forn2017ultrastrong}%
  \BibitemOpen
  \bibfield  {author} {\bibinfo {author} {\bibfnamefont {P.}~\bibnamefont
  {Forn-D{\'\i}az}}, \bibinfo {author} {\bibfnamefont {J.~J.}\ \bibnamefont
  {Garc{\'\i}a-Ripoll}}, \bibinfo {author} {\bibfnamefont {B.}~\bibnamefont
  {Peropadre}}, \bibinfo {author} {\bibfnamefont {J.-L.}\ \bibnamefont
  {Orgiazzi}}, \bibinfo {author} {\bibfnamefont {M.}~\bibnamefont {Yurtalan}},
  \bibinfo {author} {\bibfnamefont {R.}~\bibnamefont {Belyansky}}, \bibinfo
  {author} {\bibfnamefont {C.~M.}\ \bibnamefont {Wilson}}, \ and\ \bibinfo
  {author} {\bibfnamefont {A.}~\bibnamefont {Lupascu}},\ }\href@noop {}
  {\bibfield  {journal} {\bibinfo  {journal} {Nature Physics}\ }\textbf
  {\bibinfo {volume} {13}},\ \bibinfo {pages} {39} (\bibinfo {year}
  {2017})}\BibitemShut {NoStop}%
\bibitem [{\citenamefont {Breuer}\ and\ \citenamefont
  {Petruccione}(2002)}]{breuer2002theory}%
  \BibitemOpen
  \bibfield  {author} {\bibinfo {author} {\bibfnamefont {H.-P.}\ \bibnamefont
  {Breuer}}\ and\ \bibinfo {author} {\bibfnamefont {F.}~\bibnamefont
  {Petruccione}},\ }\href@noop {} {\emph {\bibinfo {title} {The theory of open
  quantum systems}}}\ (\bibinfo  {publisher} {Oxford University Press on
  Demand},\ \bibinfo {year} {2002})\BibitemShut {NoStop}%
\bibitem [{\citenamefont {Lupa{\c{s}}cu}\ \emph {et~al.}(2007)\citenamefont
  {Lupa{\c{s}}cu}, \citenamefont {Saito}, \citenamefont {Picot}, \citenamefont
  {De~Groot}, \citenamefont {Harmans},\ and\ \citenamefont
  {Mooij}}]{lupacscu2007quantum}%
  \BibitemOpen
  \bibfield  {author} {\bibinfo {author} {\bibfnamefont {A.}~\bibnamefont
  {Lupa{\c{s}}cu}}, \bibinfo {author} {\bibfnamefont {S.}~\bibnamefont
  {Saito}}, \bibinfo {author} {\bibfnamefont {T.}~\bibnamefont {Picot}},
  \bibinfo {author} {\bibfnamefont {P.}~\bibnamefont {De~Groot}}, \bibinfo
  {author} {\bibfnamefont {C.}~\bibnamefont {Harmans}}, \ and\ \bibinfo
  {author} {\bibfnamefont {J.}~\bibnamefont {Mooij}},\ }\href@noop {}
  {\bibfield  {journal} {\bibinfo  {journal} {nature physics}\ }\textbf
  {\bibinfo {volume} {3}},\ \bibinfo {pages} {119} (\bibinfo {year}
  {2007})}\BibitemShut {NoStop}%
\bibitem [{\citenamefont {Galperin}\ \emph {et~al.}(2007)\citenamefont
  {Galperin}, \citenamefont {Altshuler}, \citenamefont {Bergli}, \citenamefont
  {Shantsev},\ and\ \citenamefont {Vinokur}}]{PhysRevB.76.064531}%
  \BibitemOpen
  \bibfield  {author} {\bibinfo {author} {\bibfnamefont {Y.~M.}\ \bibnamefont
  {Galperin}}, \bibinfo {author} {\bibfnamefont {B.~L.}\ \bibnamefont
  {Altshuler}}, \bibinfo {author} {\bibfnamefont {J.}~\bibnamefont {Bergli}},
  \bibinfo {author} {\bibfnamefont {D.}~\bibnamefont {Shantsev}}, \ and\
  \bibinfo {author} {\bibfnamefont {V.}~\bibnamefont {Vinokur}},\ }\href
  {\doibase 10.1103/PhysRevB.76.064531} {\bibfield  {journal} {\bibinfo
  {journal} {Phys. Rev. B}\ }\textbf {\bibinfo {volume} {76}},\ \bibinfo
  {pages} {064531} (\bibinfo {year} {2007})}\BibitemShut {NoStop}%
\bibitem [{\citenamefont {Koch}\ \emph {et~al.}(2007)\citenamefont {Koch},
  \citenamefont {DiVincenzo},\ and\ \citenamefont
  {Clarke}}]{PhysRevLett.98.267003}%
  \BibitemOpen
  \bibfield  {author} {\bibinfo {author} {\bibfnamefont {R.~H.}\ \bibnamefont
  {Koch}}, \bibinfo {author} {\bibfnamefont {D.~P.}\ \bibnamefont
  {DiVincenzo}}, \ and\ \bibinfo {author} {\bibfnamefont {J.}~\bibnamefont
  {Clarke}},\ }\href {\doibase 10.1103/PhysRevLett.98.267003} {\bibfield
  {journal} {\bibinfo  {journal} {Phys. Rev. Lett.}\ }\textbf {\bibinfo
  {volume} {98}},\ \bibinfo {pages} {267003} (\bibinfo {year}
  {2007})}\BibitemShut {NoStop}%
\bibitem [{\citenamefont {Yoshihara}\ \emph {et~al.}(2006)\citenamefont
  {Yoshihara}, \citenamefont {Harrabi}, \citenamefont {Niskanen}, \citenamefont
  {Nakamura},\ and\ \citenamefont {Tsai}}]{PhysRevLett.97.167001}%
  \BibitemOpen
  \bibfield  {author} {\bibinfo {author} {\bibfnamefont {F.}~\bibnamefont
  {Yoshihara}}, \bibinfo {author} {\bibfnamefont {K.}~\bibnamefont {Harrabi}},
  \bibinfo {author} {\bibfnamefont {A.~O.}\ \bibnamefont {Niskanen}}, \bibinfo
  {author} {\bibfnamefont {Y.}~\bibnamefont {Nakamura}}, \ and\ \bibinfo
  {author} {\bibfnamefont {J.~S.}\ \bibnamefont {Tsai}},\ }\href {\doibase
  10.1103/PhysRevLett.97.167001} {\bibfield  {journal} {\bibinfo  {journal}
  {Phys. Rev. Lett.}\ }\textbf {\bibinfo {volume} {97}},\ \bibinfo {pages}
  {167001} (\bibinfo {year} {2006})}\BibitemShut {NoStop}%
\bibitem [{\citenamefont {Irish}\ \emph {et~al.}(2005)\citenamefont {Irish},
  \citenamefont {Gea-Banacloche}, \citenamefont {Martin},\ and\ \citenamefont
  {Schwab}}]{irish2005dynamics}%
  \BibitemOpen
  \bibfield  {author} {\bibinfo {author} {\bibfnamefont {E.}~\bibnamefont
  {Irish}}, \bibinfo {author} {\bibfnamefont {J.}~\bibnamefont
  {Gea-Banacloche}}, \bibinfo {author} {\bibfnamefont {I.}~\bibnamefont
  {Martin}}, \ and\ \bibinfo {author} {\bibfnamefont {K.}~\bibnamefont
  {Schwab}},\ }\href@noop {} {\bibfield  {journal} {\bibinfo  {journal}
  {Physical Review B}\ }\textbf {\bibinfo {volume} {72}},\ \bibinfo {pages}
  {195410} (\bibinfo {year} {2005})}\BibitemShut {NoStop}%
\bibitem [{\citenamefont {Bialczak}\ \emph {et~al.}(2007)\citenamefont
  {Bialczak}, \citenamefont {McDermott}, \citenamefont {Ansmann}, \citenamefont
  {Hofheinz}, \citenamefont {Katz}, \citenamefont {Lucero}, \citenamefont
  {Neeley}, \citenamefont {O'Connell}, \citenamefont {Wang}, \citenamefont
  {Cleland},\ and\ \citenamefont {Martinis}}]{PhysRevLett.99.187006}%
  \BibitemOpen
  \bibfield  {author} {\bibinfo {author} {\bibfnamefont {R.~C.}\ \bibnamefont
  {Bialczak}}, \bibinfo {author} {\bibfnamefont {R.}~\bibnamefont {McDermott}},
  \bibinfo {author} {\bibfnamefont {M.}~\bibnamefont {Ansmann}}, \bibinfo
  {author} {\bibfnamefont {M.}~\bibnamefont {Hofheinz}}, \bibinfo {author}
  {\bibfnamefont {N.}~\bibnamefont {Katz}}, \bibinfo {author} {\bibfnamefont
  {E.}~\bibnamefont {Lucero}}, \bibinfo {author} {\bibfnamefont
  {M.}~\bibnamefont {Neeley}}, \bibinfo {author} {\bibfnamefont {A.~D.}\
  \bibnamefont {O'Connell}}, \bibinfo {author} {\bibfnamefont {H.}~\bibnamefont
  {Wang}}, \bibinfo {author} {\bibfnamefont {A.~N.}\ \bibnamefont {Cleland}}, \
  and\ \bibinfo {author} {\bibfnamefont {J.~M.}\ \bibnamefont {Martinis}},\
  }\href {\doibase 10.1103/PhysRevLett.99.187006} {\bibfield  {journal}
  {\bibinfo  {journal} {Phys. Rev. Lett.}\ }\textbf {\bibinfo {volume} {99}},\
  \bibinfo {pages} {187006} (\bibinfo {year} {2007})}\BibitemShut {NoStop}%
\end{thebibliography}%
\bibliographystyle{apsrev4-1}

\end{document}